\documentclass[12pt]{elsart}
\usepackage[american]{babel}
\usepackage{amsbsy}     
\usepackage{amsmath}    
\usepackage{epsfig}     
\usepackage{psfig}	    

\def\gev{\,{\rm GeV}}

\def\Mpc{\,{\rm Mpc}}

\def\ev{{\,\rm eV}}

\def\cmm2{{\,\rm cm^{-2}}}
\def\cm2{{\,{\rm cm}^2}}
\def\cmm3{{\,{\rm cm}^{-3}}}
\def\gcmm3{{\,{\rm g\,cm^{-3}}}}

\def\fun#1#2{\lower3.6pt\vbox{\baselineskip0pt\lineskip.9pt
  \ialign{$\mathsurround=0pt#1\hfil##\hfil$\crcr#2\crcr\sim\crcr}}}
\def\ltsima{$\; \buildrel < \over \sim \;$}
\def\simlt{\lower.5ex\hbox{\ltsima}}
\def\gtsima{$\; \buildrel > \over \sim \;$}
\def\simgt{\lower.5ex\hbox{\gtsima}}

\newcommand{\dd}[2]{\frac{{\rm d}#1}{{\rm d}#2}}

\newcommand{\bgi}{\begin{itemize}}
\newcommand{\edi}{\end{itemize}}

\newcommand{\be}{\begin{equation}}
\newcommand{\ee}{\end{equation}}

\newcommand{\bea}{\begin{eqnarray}}                  %
\newcommand{\eea}{\end{eqnarray}}                    %

\newcommand{\beaa}{\begin{eqnarray*}}                %
\newcommand{\eeaa}{\end{eqnarray*}}                  %

\newcommand{\bgd}{\begin{description}}
\newcommand{\edd}{\end{description}}

\newcommand{\bgf}{\begin{figure}}
\newcommand{\edf}{\end{figure}}

\def\braket#1{\mathinner{\langle{#1}\rangle}}

{\catcode`\|=\active
  \gdef\Braket#1{\left<\mathcode`\|"8000\let|\bravert {#1}\right>}}
\def\bravert{\egroup\,\vrule\,\bgroup}

\def\jnl@style{}
\def\aaref@jnl#1{{\jnl@style#1}}

\def\aaref@jnl#1{{\jnl@style#1}}

\def\aj{\aaref@jnl{AJ }}                   
\def\araa{\aaref@jnl{ARA\&A }}             
\def\apj{\aaref@jnl{ApJ }}                 
\def\apjl{\aaref@jnl{ApJ }}                
\def\apjs{\aaref@jnl{ApJS }}               
\def\ao{\aaref@jnl{Appl.~Opt. }}           
\def\apss{\aaref@jnl{Ap\&SS }}             
\def\aap{\aaref@jnl{A\&A }}                
\def\aapr{\aaref@jnl{A\&A~Rev. }}          
\def\aaps{\aaref@jnl{A\&AS }}              
\def\azh{\aaref@jnl{AZh }}                 
\def\baas{\aaref@jnl{BAAS }}               
\def\jrasc{\aaref@jnl{JRASC }}             
\def\memras{\aaref@jnl{MmRAS }}            
\def\mnras{\aaref@jnl{MNRAS }}             
\def\pra{\aaref@jnl{Phys.~Rev.~A }}        
\def\prb{\aaref@jnl{Phys.~Rev.~B }}        
\def\prc{\aaref@jnl{Phys.~Rev.~C }}        
\def\prd{\aaref@jnl{Phys.~Rev.~D }}        
\def\pre{\aaref@jnl{Phys.~Rev.~E }}        
\def\prl{\aaref@jnl{Phys.~Rev.~Lett. }}    
\def\pasp{\aaref@jnl{PASP }}               
\def\pasj{\aaref@jnl{PASJ }}               
\def\qjras{\aaref@jnl{QJRAS }}             
\def\skytel{\aaref@jnl{S\&T }}             
\def\solphys{\aaref@jnl{Sol.~Phys. }}      
\def\sovast{\aaref@jnl{Soviet~Ast. }}      
\def\ssr{\aaref@jnl{Space~Sci.~Rev. }}     
\def\zap{\aaref@jnl{ZAp }}                 
\def\nat{\aaref@jnl{Nature }}              
\def\iaucirc{\aaref@jnl{IAU~Circ. }}       
\def\aplett{\aaref@jnl{Astrophys.~Lett. }} 
\def\apspr{\aaref@jnl{Astrophys.~Space~Phys.~Res. }}
\def\bain{\aaref@jnl{Bull.~Astron.~Inst.~Netherlands }} 
\def\fcp{\aaref@jnl{Fund.~Cosmic~Phys. }}  
\def\gca{\aaref@jnl{Geochim.~Cosmochim.~Acta }}   
\def\grl{\aaref@jnl{Geophys.~Res.~Lett. }} 
\def\jcp{\aaref@jnl{J.~Chem.~Phys. }}      
\def\jgr{\aaref@jnl{J.~Geophys.~Res. }}    
\def\jqsrt{\aaref@jnl{J.~Quant.~Spec.~Radiat.~Transf. }}
\def\memsai{\aaref@jnl{Mem.~Soc.~Astron.~Italiana }}
\def\nphysa{\aaref@jnl{Nucl.~Phys.~A }}   
\def\physrep{\aaref@jnl{Phys.~Rep. }}   
\def\physscr{\aaref@jnl{Phys.~Scr }}   
\def\planss{\aaref@jnl{Planet.~Space~Sci. }}   
\def\procspie{\aaref@jnl{Proc.~SPIE }}   

\begin{document}

\begin{frontmatter}

\title{High energy gamma ray counterparts of astrophysical sources 
of ultra-high energy cosmic rays}

\author[iasf,palermo]{Carlo Ferrigno\thanksref{corr1}}
\author[inaf,infnfi]{Pasquale Blasi\thanksref{corr2}}
\author[genova,infnge]{Daniel De Marco\thanksref{corr3}}

\address[iasf]{Istituto di Astrofisica Spaziale e Fisica Cosmica - CNR, \\
Via Ugo La Malfa 153 - 90146 Palermo - ITALY}
\address[palermo]{Universit\`a degli Studi di Palermo\\
Via Archirafi, 36 - 90123 Palermo, ITALY}

\address[inaf]{INAF/Osservatorio Astrofisico di Arcetri,\\
Largo E. Fermi, 5 - 50125 Firenze, ITALY}
\address[infnfi]{INFN/Sezione di Firenze, ITALY}

\address[genova]{Universit\`a degli Studi di Genova,\\
Via Dodecaneso, 33 - 16146 Genova, ITALY}
\address[infnge]{INFN/Sezione di Genova, ITALY}

\thanks[corr1]{E-mail: ferrigno@pa.iasf.cnr.it}
\thanks[corr2]{E-mail: blasi@arcetri.astro.it}
\thanks[corr3]{E-mail: ddm@ge.infn.it}

\begin{abstract}
If ultra-high energy cosmic rays (UHECRs) are accelerated at astrophysical 
point sources, the identification of such sources can be achieved if there 
is some kind of radiation at observable wavelengths that may be 
associated with the acceleration and/or propagation processes. No 
radiation of this type has so far been detected or at least no such 
connection has been claimed. The process of photopion production during
the propagation of UHECRs from the sources to the Earth results in the
generation of charged and neutral pions. The neutral (charged) pions in 
turn decay to gamma quanta and electrons that initiate an electromagnetic 
cascade in the universal photon background. 
We calculate the flux of this gamma radiation in the
GeV-TeV energy range and find that for source luminosities compatible with 
those expected from small scale anisotropies in the directions of arrival
of UHECRs, the fluxes can be detectable by future Cerenkov gamma ray 
telescopes, such as VERITAS and HESS, provided the intergalactic 
magnetic field is not larger than $\sim 10^{-10}$ Gauss and for source
distances comparable with the loss length for photopion production.
\end{abstract}
\end{frontmatter}

\section{Introduction}

The discovery of the cosmic microwave background radiation (CMBR) 
\cite{penzias} has been a major breakthrough not only for cosmology but
for cosmic ray physics as well. Soon after its discovery, it was shown 
that cosmic rays would suffer severe energy losses due to the inelastic
process of photopion production, provided the energy of cosmic rays 
being higher than the kinematic threshold for this process. This effect
was predicted to result in a flux suppression at energies around $10^{20}$~eV
\cite{greisen66,zk66} 
for a homogeneous distribution of the sources, something
which is now known as the GZK feature. Despite the efforts dedicated
to the experimental detection of this feature, to date we do not have
a definite answer to whether the GZK suppression is there or not. The
two largest detectors currently operating, AGASA and HiRes, give results 
which appear discrepant, although the statistics of events does not allow 
us to infer a definitive conclusion on this issue \cite{demarco1}. 
The detection of the GZK feature in the spectrum of UHECRs would be the 
clearest proof that UHECRs are generated at extragalactic sources. 

Despite the efforts in building several large experiments for the
detection of particles with the highest energies in the cosmic ray
spectrum, the nature of the sources of these particles and the
acceleration processes at work are still unknown. No trivial counterpart
has been identified by any of the current experiments: there is no
significant association with any large scale local structure nor with
single sources. The lack of counterparts is particularly puzzling at
the highest energies, where the loss length of the detected particles
is small enough that only a few sources can be expected to be located
within the error box of current experiments. 

It has been recently proposed that a statistically significant
correlation may exist between the arrival directions of UHECRs with
energy above $2\times 10^{19}$ eV and the spatial location of BL Lac objects,
with redshifts larger than 0.1 \cite{tkachev1,tkachev2} (but see also 
\cite{efs} for a criticism and \cite{ttreply} for the reply). Since at these 
energies the loss length is comparable with the size of the universe, no 
New Physics would be needed to explain the possible correlation. On the other
hand, since no BL Lac is known to be located close to the Earth, the 
spectrum of UHECRs expected from these sources would have a very pronounced
GZK cutoff. 

The potential for discovery of the sources has recently improved after 
the identification of a few doublets and triplets of events clustered on 
angular scales comparable with the experimental angular resolution of 
AGASA. A recent analysis of the combined results of most UHECR experiments 
\cite{uchihori} revealed 8 doublets and two triplets on a total of 92 events 
above $4\times 10^{19}$ eV (47 of which are from AGASA). The recent HiRes
data do not show evidence for significant small angle clustering, but this
might well be the consequence of the smaller statistics of events and of 
the energy dependent acceptance of this experiment: one should remember 
that in order to reconstruct the spectrum of cosmic rays, namely to account 
for the unobserved events, a substantial correction for this energy dependence 
need to be carried out in HiRes (the acceptance is instead a flat function
of energy for AGASA at energies above $10^{19}$ eV). This correction however 
does not give any information on the distribution of arrival directions 
of the {\it missed} events.

The statistical significance of these small scale multiplets of events 
have recently been questioned in \cite{finley2003}. 
If the appearance of these multiplets in the data will be confirmed by future 
experiments as not just the result of a statistical fluctuation or focusing in 
the galactic magnetic field \cite{harari}, then the only way to explain their 
appearance is by assuming that the sources of UHECRs are in fact point sources.
This would represent the first true indication in favor of astrophysical 
sources of UHECRs, since the clustering of events in most top-down scenarios 
for UHECRs seems unlikely. A clear identification of the GZK feature in the 
spectrum of UHECRs would make the evidence in favor of astrophysical sources 
even stronger.

In some recent work \cite{blasi2003,will,tom,dubovsky,fodor} it has been 
shown that the small scale anisotropies can be used as a powerful tool to 
measure the number density of sources of UHECRs. In particular, in 
\cite{blasi2003} the number density of sources that best fits the 
observed number of doublets and triplets was found to be of order
$10^{-6}-10^{-5}$ $\rm Mpc^{-3}$, with an average luminosity per source
of $2\times 10^{42}-2\times 10^{43}\rm erg~s^{-1}$ at energies in excess
of $10^{19}$ eV. The extrapolation of these numbers down to GeV energies 
are strongly dependent upon the injection spectrum, assumed here in 
such a way to fit the observed spectrum of cosmic rays above $10^{19}$ eV.

Alternative estimates of the density of sources of UHECRs recently appeared
in the literature: the number obtained in \cite{yoshi} is in fair 
agreement with the result of \cite{blasi2003}, while more complex is
the comparison with the case investigated in \cite{sigl1,ensslin}. In 
\cite{sigl1}, a source density of $\sim 10^{-4}-10^{-3} \rm Mpc^{-3}$ was 
derived, while the large scale isotropy of UHECRs was used to infer a local 
field strength of $\sim 0.1 \mu G$ in the local supercluster. This conclusion 
was mainly due to the fact that the authors neglected the contribution of 
distant sources of cosmic rays with energy above $4\times 10^{19}$ eV. 
When this effect is accounted for, as in a more recent paper \cite{ensslin} 
by the same authors, the revised estimate of the source density appears to 
be in rough agreement with that of \cite{blasi2003} for the cases of weak 
magnetization. The authors of \cite{ensslin}
also investigate the cosmic ray propagation in a magnetized universe,
as obtained from numerical simulations of large scale structure
formation with passive magnetic fields. In these cases the estimate
of the source density becomes more model dependent.
It is probably worth stressing however, that the magnetic fields 
obtained in \cite{ensslin} do not appear to be compatible with those derived
in the numerical MHD simulations of \cite{grasso}: while the magnetic field 
obtained inside clusters of galaxies in both numerical approaches appears to 
have roughly the same strength, the fields outside clusters differ by 2-4 
orders of magnitude. Not surprisingly the conclusions that the two groups 
draw on the propagation of UHECRs are incompatible with each other. It would 
be auspicable that the two pieces of work can achieve some kind of agreement, 
so to clarify the situation. 

The interactions of cosmic ray protons at the highest energies with the 
photons in the CMBR, responsible for the appearance of the GZK feature
in the diffuse spectrum of UHECRs, also generate gamma radiation through 
the decay of neutral pions and electrons through the decay of charged
pions (actually mostly positrons are produced, but hereafter we will 
refer to both electrons and positrons as {\it electrons}). The very 
high energy gamma radiation produced 
in this way initiates an electromagnetic cascade that results in the 
appearance of a gamma ray flux in the energy region accessible to 
Cerenkov detectors. We investigate in detail this process and assess
the detectability of this signal from the directions where the sources 
of UHECRs are located.

The paper is organized as follows: in section \ref{sec:cascade} we present 
an analytical estimate of the development of the electromagnetic cascade 
initiated by UHECRs; in section \ref{sec:background} we detail the background 
radiation field we used for our calculations; in section 
\ref{sec:detailedcascade} we present the detailed calculation of the 
electromagnetic cascade developement; 
in section \ref{sec:results} we present our results for different luminosities
of the sources and different values and topologies of the extragalactic 
magnetic field; in section \ref{sec:conclusions} we draw our conclusions.

\section{The electromagnetic cascade initiated by UHECRs: an analytical
approach}\label{sec:cascade}

The purpose of this section is to provide an estimate of the expected 
flux of gamma rays due to the cascade initiated by UHECRs. A detailed
calculation will be presented in the next sections, where many effects
neglected here will be properly accounted for.

UHECRs with energy above the threshold for photopion production are
kinematically allowed to interact inelastically with the photons in
the CMBR and generate charged and neutral pions mainly through the 
reactions
\be
p + \gamma \to \pi^+ + n ~~~~~~~~~ p + \gamma \to \pi^0 + p.
\label{eq:photopion}
\ee
The decay of neutral pions results almost instantaneously in the 
production of two photons with typical energy $E_\gamma\sim E_\pi/2\sim 
0.1 E_p$ (here, for the purpose of a rough estimate we assumed an
inelasticity of $\sim 20\%$). 

These photons start an electromagnetic cascade through pair
production on the universal photon background. A similar cascade is
initiated by positrons resulting from the decays of $\pi^+$'s. 

Cosmic rays with energy above $\sim 4\times 10^{17}$ eV can 
interact with the CMBR through proton pair-production, which results 
in electrons and positrons which can also initiate an electromagnetic 
cascade. This process becomes however relevant for the propagation of
UHECRs only at energies above $\sim 10^{18}$ eV, when adiabatic energy
losses due to the expansion of the universe become less severe. 
We estimated that the contribution of the proton pair production to
the electromagnetic cascade is of few percent of the contribution initiated
through photopion production but can be up to $\sim 50\%$ for distant 
sources. In the rest of this calculation we shall focus on the contribution 
from photopion production, while a full treatment of additional contributions 
will be properly considered in an upcoming paper (Ferrigno, Blasi \& De Marco, 
in preparation). 

We will provide a detailed description of our calculations of the development 
of the cascade in the next section. Here we estimate the expected effect 
using an approach which was proposed in \cite{smirnov} and detailed in 
\cite{bible}. 

Let us consider a simple situation in which the cosmic photon background
is dominated by the radio radiation and the CMBR. For our purposes here
we can assume that the typical energy of the radio photons is 
$\epsilon_{radio}=4\times 10^{-9}$ eV (corresponding to a frequency 
of $1$ MHz). For the CMBR, the reference energy will be taken as 
$\epsilon_{CMB}=6\times 10^{-4}$ eV. 

Most of the gamma rays that initiate the electromagnetic cascade have
energy $E_\gamma\ge 10^{19}$ eV. We define here the two quantities
$x_{radio}=E \epsilon_{radio}/m_e^2$ and $x_{CMB}=E \epsilon_{CMB}/m_e^2$,
where $E$ is the energy of the leading particle, which in our case starts
as being either a photon or an electron. 
Clearly $x_{radio}\ll x_{CMB}$ therefore the cascade develops
first on the radio photon background. For the particles that start the
cascade, $x_{radio}\approx 0.1-1$. In fact $x_{radio}$ can be quite higher
for higher energy photons, that are also generated as a result of the
photopion production. For the range of values $x_{radio}\gg 1$ the pair
production of photons and the inverse Compton scattering (ICS) of electrons
once they have been produced, occur in the Klein-Nishina (KN) limit and the 
leading particle loses a fraction of its energy which is approximately
$f=\Delta E/E \approx 1/\ln(2 x_{radio})$. The energy of the secondary 
particle produced will therefore be $E'\approx f E$. The KN regime
stops however when $x_{radio}\sim 1$, which is almost immediately the
case for the interaction on the radio background. At this point the 
cascade develops efficiently through pair production and ICS down to
the point when photon energies fall below the threshold for pair 
production, $E_\gamma^{th,radio}=m_e^2/\epsilon_{radio}$, and the number of 
electrons remains constant at later times. Therefore we can write for
the electron spectrum
\be
q_e(E_e) = q_0 ~~~ \rm if~~~E_e < E_\gamma^{th,radio}/2.
\ee
At higher energies, conservation of energy gives an electron spectrum 
that declines as the first power of the electron energy 
\cite{smirnov,bible}. The photon spectrum in the cascade is generated
through the ICS of these electrons. 
Since the ICS occurs in the Thomson regime, it is simple to demonstrate
that the electrons with a flat spectrum ($q_e(E_e)=q_0$) generate a
gamma ray spectrum $n_\gamma(E_\gamma)\propto E_\gamma^{-3/2}$ up
to a maximum energy
\be 
E_X^{radio} = \frac{4}{3} \left(\frac{E_\gamma^{th,radio}}{2 m_e}\right)^2 
\epsilon_{radio} = \frac{1}{3} \frac{m_e^2}{\epsilon_{radio}}.
\ee
The electrons with spectrum $q_e(E_e)\propto E_e^{-1}$ generate inverse
compton photons with spectrum $n_\gamma(E_\gamma)\propto E_\gamma^{-2}$
up to the energy $E_\gamma^{max,radio} = m_e^2/\epsilon_{radio}$. At higher
energies there are no gamma rays because the electron spectrum is cut off.
The normalization in the gamma ray spectrum is such that the total energy
in it equals the energy in the primary particles (gamma rays and/or 
electrons). 

It is worth noticing that the spectrum of the initial gamma rays and 
electrons does not explicitely enter the calculation: the spectrum of 
the final cascade is universal. In other words the electromagnetic cascade 
behaves as a sort of calorimeter that redistributes the initial energy 
into gamma rays with a given spectrum. Violations to this rule will be 
seen and discribed in the next section, but the basic result is not 
substantially changed.

Once this first part of the cascade has developed, the gamma rays generated
so far can produce a second cascade on the CMB photon background. This 
second cascade follows exactly the same steps described above. We introduce
here a new energy scale 
\be 
E_X^{CMB} = \frac{4}{3} \left(\frac{E_\gamma^{th,CMB}}{2 m_e}\right)^2 
\epsilon_{CMB} = \frac{1}{3} \frac{m_e^2}{\epsilon_{CMB}}.
\ee
The electron spectrum is flat up to the energy $E_\gamma^{th,CMB}/2$
as in the first part of the cascade. The parent gamma ray spectrum, as 
we found above, has the first change in slope at the energy $E_X^{radio}$, 
therefore it follows that a slope change will appear in the electron spectrum 
generated through pair production, at the energy $E_X^{radio}/2$. At energies
$E_\gamma^{th,CMB}/2\leq E_e \leq E_X^{radio}/2$ the electron spectrum
is $q_e(E_e)\propto E_e^{-1/2}$, while at higher energies it again steepens
to $q_e(E_e)\propto E_e^{-1}$. The final gamma ray spectrum can be 
calculated as ICS of the electrons, and after simple calculations we
obtain:
\be
n_\gamma(E_\gamma) = n_0 \times
\begin{cases}
\left(\frac{E_\gamma}{E_X^{CMB}}\right)^{-3/2} ~~~ E_\gamma \leq E_X^{CMB}\\
\left(\frac{E_\gamma}{E_X^{CMB}}\right)^{-7/4} ~~~ E_X^{CMB}\leq E_\gamma \leq
\frac{1}{3} \frac{E_X^{radio}}{m_e^2}\epsilon_{radio}=\tilde E \\
\left(\frac{E_X^{radio~2}\epsilon_{radio}}{3 E_X^{CMB} m_e^2}\right)^{-7/4}
\left(\frac{E_\gamma}{\tilde E}\right)^{-2} ~~~ \tilde E\leq E_\gamma\leq
E_\gamma^{th,CMB}\\
0 ~~~ E_\gamma>E_\gamma^{th,CMB}.\\
\end{cases}
\ee
The constant $n_0$ should be calculated by requiring that the total energy
of the primary gamma rays and electrons is converted into the energy of 
the cascade. The calculations outlined above assume in principle that we 
start with a gamma ray and/or electron beam at some distance. 
In fact, for the case of gamma rays generated from the decay
of neutral pions created in turn in the $p+\gamma$ inelastic interactions 
this is not the case: gamma rays and electrons are generated at any distance 
from the source as long as the energy of the primary cosmic rays remains 
in excess of the threshold for photopion production. This means that
most gamma rays are injected in the first few interaction lengths from
the source, and their cascade starts at the production point. It is
therefore clear that the result provided by the analytical calculation
should be taken only as the estimate of an order of magnitude of the 
expected effect.
Another complication consists in the lack of complete development of 
the cascade for those parts of the cascade initiated too close to
the Earth. All these effects, included in our detailed calculations, 
tend to decrease the flux of lower energy gamma rays compared with 
the rough estimate that we report in this section. On the other hand,
while this estimate ignores the infrared background, some energy is 
transferred to the cascade due to the interaction with this third photon 
background. 

In order to have some reference numbers, we assume to have a source 
of UHECRs with a luminosity $L_{CR}$, in the form of cosmic rays 
with energy above some value, that we take here to be $10^{19}$
eV. Not all of this luminosity can be converted into gamma rays because
of the kinematic threshold in the photopion production process. If the
injection spectrum is taken as $E^{-2.6}$ \cite{demarco1}, the energy 
that can potentially be converted to gamma rays is about $20\% L_{CR}$.

Inserting numbers in the previous expressions for the cascade spectrum,
we obtain the following estimate of the flux in the energy region below
$E_X^{CMB}=140~\rm TeV$:
\be 
F(E_\gamma) \approx 7.4\times 10^{-14}
\left(\frac{L_{CR}}{10^{43}\rm erg~s^{-1}}\right) 
\left(\frac{d}{10\rm Mpc}\right)^{-2}
\left(\frac{E_\gamma}{140\rm TeV}\right)^{-1/2}~\rm photons~cm^{-2}~s^{-1}
\ee
In the TeV region, the expected fluxes are comparable with the sensitivities
of upcoming Cerenkov gamma ray experiments such as VERITAS and HESS.

\section{The background radiation field}
\label{sec:background}

In this section we describe the photon background adopted in our calculations.
For the sake of clarity we discuss separately the cases of the radio, 
microwave and IR-Optical-UV parts of the spectrum. It is worth stressing 
the intrinsic differences between the CMBR on one side and the other 
backgrounds. The former is of cosmological origin and is 
pretty well known to have a Planckian spectrum with temperature at the
present cosmic time of $T=2.728\pm 0.004 $~K \cite{COBE}. The radio and 
UV-Optical-IR backgrounds are instead due to emission processes in 
astrophysical objects and are much more poorly known observationally. 
In particular the knowledge
of the radio background in the frequency range below a few MHz is very 
poor: our Galaxy is opaque to these radio waves due to free-free
absorption, so that we cannot measure directly the universal photon 
background at these frequencies. We are therefore forced to rely only upon
theoretical calculations and a few observations at higher frequencies. 
The UV-Optical-IR background is also known only within a factor of a few,
due to the difficulty in separating the extragalactic contribution from 
the galactic emission \cite{elbaz2002}.

{\it The UV-Optical-IR background}

In a very general way, the fluence $s_\nu$ (in units of energy per unit time 
per unit surface) in photons of frequency $\nu$ contributed by sources with
luminosity function per comoving volume $\phi(L,z)$ can be written as
follows \cite{peacock}:
\be
s_\nu= c \int_0^{\infty}\mathrm{d}z \left|\dd{t}{z}\right|
\int_{L_{min}}^{L_{max}}\mathrm{d}L\phi(L,z)f_\nu(z,L),
\label{eq:fluence}
\ee
where $f_\nu(z,L)$ is the emissivity of the sources with luminosity $L$ at 
red-shift $z$.

The UV-Optical-IR background is thought to be generated by galaxies, 
therefore we calculate here such background as a function of redshift 
following the procedure detailed in \cite{chary2001}. In the calculations,
we adopt a cosmology with $H_0=71$~km/s/Mpc, $\Omega_m h^2=0.135$ with 
$h=H_0/100$ and $\Omega=1$ (thus $\Omega_\Lambda = 0.73$) as derived by 
WMAP \cite{wmap}.

The procedure described in \cite{chary2001} is based on building a set of 
synthetic galaxy spectral energy distributions (SEDs) as functions of the 
luminosity in the ISO band LW3 centered at $15\;\mu m$. These spectra are
built as linear combinations of the synthetic spectra of four galaxies,
Arp~220, NGC~6090, M~82 and M~51, as prototypes of ultra luminous infrared, 
luminous infrared, starburst and normal galaxies respectively,
as obtained from the online library of synthetic spectra created with 
the GRASIL code \cite{grasil}. In this way we compute the emissivity 
$f_\nu(z,L)$ at $z=0$. The red-shifted emissivity is derived by scaling 
the luminosity by the function $g(z)$ of equations (\ref{eq:g1}) and 
(\ref{eq:g2}) so that:
\be
f_\nu(z,L)=f_\nu(0,L/g(z)).
\ee

We divide the luminosity interval of the $15 \; \mu \mathrm{m}$ luminosity 
function of \cite{xu2000} in 30 bins and for each bin we take the SED computed 
with the linear combination of the synthetic spectra. We use the best fit 
combination of luminosity and density evolution computed by \cite{chary2001} 
so that the luminosity function reads
\be
\phi(L, z)=n(z) \phi\left( \frac{L}{g(z)},0\right)
\ee
with 
\be
\phi(L,0)= \dd{}{L} \left[ N\left( \frac{L}{L_*}\right)^a
\left(1+\frac{L}{L_*}\right)^b \right]
\ee
with $N=10^{-0.22}\Mpc^{-3}$, $a=0.1$, $b=0.1$ \cite{xu2000,xu1998}. The 
red-shift evolution is described by the quantities
\bea
n(z)&=&\left(1+z\right)^{\alpha_1}\\
g(z)&=&\left(1+z\right)^{\beta_1}
\label{eq:g1}
\eea
for $z<z_T$ and
\bea
n(z)&=&n(z_T)\left(\frac{1+z}{1+z_T}\right)^{\alpha_2}\\
g(z)&=&g(z_T)\left(\frac{1+z}{1+z_T}\right)^{\beta_2}
\label{eq:g2}
\eea
for $z \ge z_T$. The parameters $z_T$, $\alpha_k$ and $\beta_k$ are as 
computed in \cite{chary2001}.

\begin{figure}
  \begin{center}
    \epsfig{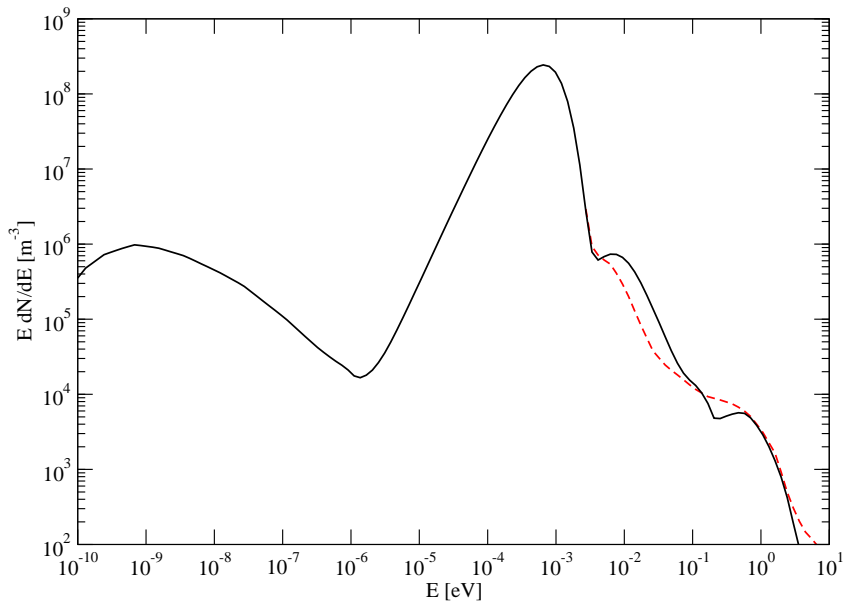}
	\caption{The local photon background we used. The main peak is 
due to the CMB, the radio background is taken from \cite{biermann96} in 
the hypothesis of red-shift evolving sources, the IR-UV background is taken 
from \cite{chary2001}. The dashed curve shows the result obtained in 
\cite{stecker}.}
\label{back}
\end{center}
\end{figure}

The IR-UV spectrum has an infrared luminosity between $3.5$ and $1000\;\mu 
\mathrm{m}$ of 34~$\mathrm{nW\, m}^{-2}\mathrm{sr}^{-1}$. The observational 
bound is $50\pm25\;\mathrm{nW\, m}^{-2}\mathrm{sr}^{-1}$ and other theoretical 
estimates lay in the range between 23 and 73~$\mathrm{nW\, m}^{-2}
\mathrm{sr}^{-1}$ (see \cite{dwek2002} and references therein). A direct
comparison of our background with that obtain in \cite{stecker} (dashed line) is 
shown in Fig.~\ref{back}.

{\it The radio background}

\begin{figure}[ht]
\begin{center}
\psfig{figure=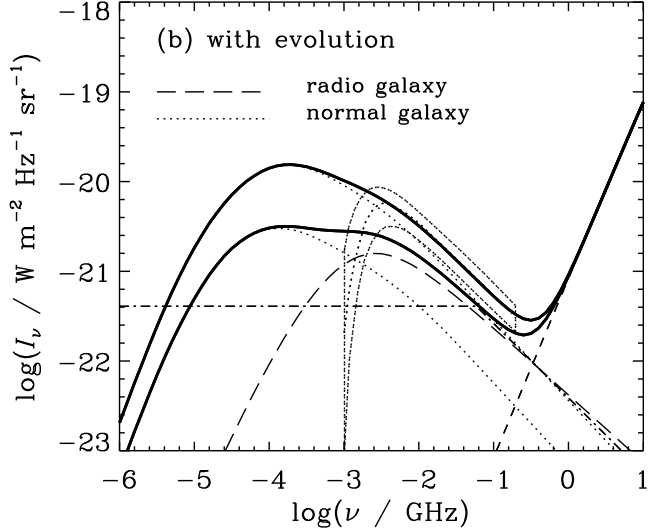, width=0.8\textwidth}
\end{center}
\caption{Figure from Ref. \cite{biermann96}. Contributions of normal 
galaxies (dotted curves), radio galaxies 
(long dashed curve), and the cosmic microwave background (short dashed curve) 
to the extragalactic radio background intensity (thick solid curves) with 
pure luminosity evolution for all sources (upper curves), and for radio 
galaxies only (lower curves). The dotted band gives an observational estimate 
of the total extragalactic radio background intensity \cite{alexander69} and 
the dot-dash curve gives an earlier theoretical estimate \cite{bere1}.} 
\label{radio}
\end{figure}

As pointed out before, the universal radio background (URB) is not very well
known mostly because it is difficult to disentangle the Galactic and 
extragalactic components and because of the free-free absorption 
in the galactic disk and halo at low frequencies. Observations have provided 
us with an estimate of the URB \cite{alexander69}, while an early theoretical 
estimate was given in \cite{bere1}. 
More recently, an attempt has been made to calculate
the contribution to the URB from radio-galaxies and AGNs~\cite{biermann96}
(see Fig.~\ref{radio}). The issue does not seem to be settled, but 
unfortunately it seems to be confined to the field of theoretical 
calculations, at least in the frequency range below $\sim 1$ MHz.

In our calculation we adopt the URB as given in \cite{biermann96} for
the case of pure luminosity evolution of normal galaxies and radio-galaxies.

The local photon background obtained as a superposition of the IR-Optical-UV, 
the radio and the microwave background is plotted in fig. \ref{back}, 
where the dashed curve shows, for comparison, the result of Ref. 
\cite{stecker}. We checked that our results do not change appreciably 
by changing the IR background, the reason being that most part of the 
cascade develops on the URB and on the CMB radiation.

\section{The Electromagnetic cascade: a detailed calculation}
\label{sec:detailedcascade}

In this section we describe our calculation of the electromagnetic
cascade initiated by a proton with ultra-high energy through photopion
production off the photons of the CMBR. In \S \ref{sec:prod} we 
describe the calculation of the rate of injection of gamma rays and
electrons in the cascade due to discrete interactions of photopion 
production. In \S \ref{sec:cas} we describe in detail the propagation 
of electrons and gamma rays in the background radiation field. 

\subsection{Gamma rays and electrons from photopion production}
\label{sec:prod}

At energies larger than the threshold for photopion production, given by
\be
E_{th} = \frac{m_\pi^2 + 2 m_\pi m_p}{4 \epsilon},
\ee
where $\epsilon$ is the energy of the target photon, a proton 
can generate a pion in the final state, in the reaction of photopion 
production as written in Eq. \ref{eq:photopion}. The decay of the 
$\pi^0$ or $\pi^+$ generates either gamma rays ($\pi^0\to\gamma\gamma$) 
or electrons ($\pi^+\to \mu^+ \nu_\mu;~\mu\to e + \nu_\mu + \nu_e$).
Both these secondary products can initiate a cascade in the universal 
photon background calculated in the previous section. 

The propagation of UHECRs from a source at a given distance is simulated
using the Montecarlo approach presented in \cite{demarco1}. However, in
the version of the simulation adopted here we improved the code by
simulating the single proton-photon interactions using SOPHIA
\cite{sophia,stanev}.  This event generator treats an interaction either 
via baryon resonance excitation, one particle t-channel exchange (direct
one-particle production), diffractive particle production and
(non-diffractive) multiparticle production using string fragmentation.
The distribution and momenta of the final state particles are calculated
from their branching ratios and interaction kinematics in the center of
mass frame, and the particle energies and angles in the lab frame are
calculated by Lorentz transformations. SOPHIA takes care also of the
decay of all unstable particles (except for neutrons) using standard
Monte Carlo methods of particle decay according to the available phase
space. The neutron decay is implemented separately due to its much
longer lifetime. The SOPHIA event generator has been shown to be in 
good agreement with available accelerator data. 
A detailed description of the code including the sampling methods,
the interaction physics used, and the performed tests can be found in
Ref. \cite{sophia}.

\begin{figure}
  \begin{center}
    \epsfig{figure=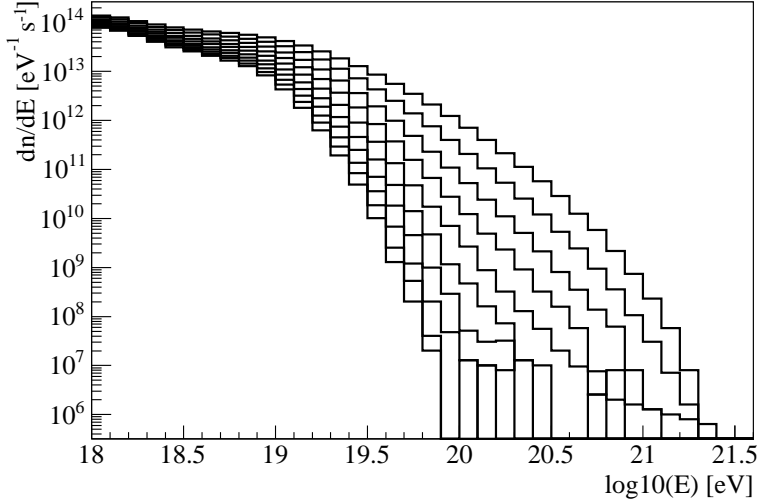,width=0.8\textwidth}
	\caption{Photon injection spectra accumulated in bins of 10~Mpc 
width from the source (from top to bottom).}
\label{fig:injgamma}
\end{center}
\end{figure}

\begin{figure}
  \begin{center}
    \epsfig{figure=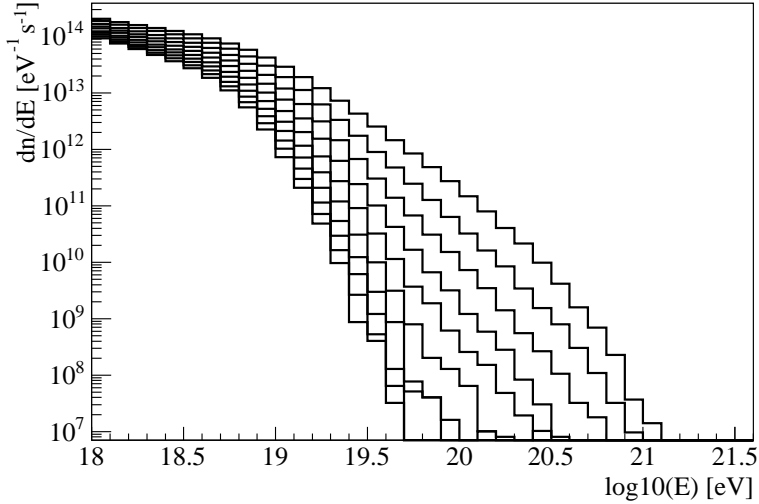,width=0.8\textwidth}
	\caption{Positron injection spectra accumulated in bins of 10~Mpc 
width from the source (from top to bottom).}
\label{fig:injelec}
\end{center}
\end{figure}

The gamma ray and positron yields used in our calculation are computed
by simply accounting for the decay of neutral and charged pions.
The spectra of gamma rays at different distances from the source,
for a given distance between the source and the observer are plotted
in Fig. \ref{fig:injgamma}. The spectra of electrons/positrons coming from the
decay of charged pions for the same distances from the source are
shown in Fig. \ref{fig:injelec}. 

As shown in \cite{demarco1}, the injection spectrum of UHECRs that 
best fits the observed data of AGASA and HiRes in the energy region 
below $10^{20}$ eV is a power law $E^{-\gamma}$ with $\gamma=2.6$ 
in the case of sources that do not have luminosity evolution with
redshift. The injection spectrum flattens to $E^{-2.4}$ for the
case of sources with strong luminosity evolution $L(z)\propto (1+z)^4$.

In \cite{blasi2003} a detailed analysis of the consequence of 
small scale anisotropies in the directions of arrival of UHECRs
was carried out. It was found there that the number of doublets
and triplets observed by AGASA suggests that the sources of UHECRs
are powerful relatively rare sources, with a typical number density
$10^{-6}-10^{-5}~\rm Mpc^{-3}$ and a luminosity at energies above
$10^{19}$ eV of $\sim 2\times 10^{42}-2\times 10^{43}~\rm erg~s^{-1}$.


The reactions of photopion production during the propagation of UHECRs
from the sources to the Earth generate at any time new gamma rays and
electrons, that start new cascades that will eventually reach the Earth.

\subsection{The development of the electromagnetic cascade}
\label{sec:cas}

As already discussed at length in section \ref{sec:cascade}, the 
electromagnetic cascade is mainly driven by Inverse Compton Scattering 
(ICS) of high energy electrons on the low energy background photons and 
Pair Production (PP) of 
high energy photons on the low energy background photons. Higher order QED 
processes become important only for energies exceeding $10^{21}\ev$ and have 
a significant impact only in the absence of magnetic fields and for a narrow 
peaked background spectrum \cite{bs2000}. In our calculations we only include
ICS, PP and synchrotron emission in the presence of an intergalactic magnetic
field.

Here we use the following notation: $f_e(E, t)$ is the differential
spectrum of electrons at time $t$ during the evolution of the cascade, 
$f_\gamma(E, t)$ is the differential spectrum of photons at the same time
$t$. We calculate these spectra by solving the coupled Boltzman equations
that describe the evolution of $f_e$ and $f_\gamma$ due to energy losses.

The Boltzmann equations read \cite{lee98}:
\begin{multline}
\dd{f_e(E,t)}{t}=-\int\mathrm{d}\epsilon n_b(\epsilon) 
R_{ICS}(E,\epsilon) f_e(E,t) + \\
\int\mathrm{d}\epsilon n_b(\epsilon)\int\mathrm{d}E' P_{ICS,e}f_e(E',t)+ 
\int\mathrm{d}\epsilon n_b(\epsilon)\int\mathrm{d}E' P_{PP}f_\gamma(E',t) + q_e(E,t)
\end{multline}
\begin{multline}
\dd{f_\gamma(E,t)}{t}=-\int\mathrm{d}\epsilon n_b(\epsilon) 
R_{PP}(E,\epsilon) f_\gamma(E,t) + \\
\int\mathrm{d}\epsilon n_b(\epsilon)\int\mathrm{d}E' 
P_{ICS,\gamma}f_e(E',t) + q_\gamma(E,t),
\label{boltzmann}
\end{multline}
where we introduced the angle averaged cross sections as
\be
R (E,\epsilon) = \int_{-1}^1 \mathrm{d}\mu \frac{1-\mu \beta_{e,\gamma}}{2}
\sigma (s(\epsilon_b,E,\mu))
\ee
and the angle averaged differential cross sections as
\be
P (E';E,\epsilon) = \int_{-1}^1 \mathrm{d}\mu \frac{1-\mu \beta_{e,\gamma}}{2}
\dd{ \sigma (s(\epsilon_b,E,\mu))}{E'}.
\ee
Here $s(\epsilon_b,E,\mu)$ is the center of mass squared energy for a 
particle with energy $E$ scattering a photon with energy $\epsilon_b$
at angle having cosine $\mu$. $n_b(\epsilon)$ is the spectrum of background 
photons. $ q_e(E,t)$ and $q_\gamma(E,t)$ are the source functions for 
electrons and photons respectively, calculated as described in 
\S \ref{sec:prod}. At any time new electrons and gamma rays are generated 
and start new cascades, due to the reactions of photopion production of 
protons.

\subsubsection{Pair production}
The pair production cross section is given by \cite{svensson1982}:
\be
\sigma_{PP}=\sigma_{T} \frac{3}{16}\left(1-\beta^2\right) \left[ 
\left( 3 - \beta^4 \right) \ln \frac{1+\beta}{1-\beta} -2 \beta 
\left( 2-\beta\right) \right],
\label{sigma_PP}
\ee
where $\beta \def \sqrt{1-4m_e^2/s}$ is the velocity of the outgoing electron 
in the Center of Mass (CM) frame. 

The threshold for pair production is:
\be
E_{th}=\frac{m_E^2}{\epsilon}\simeq 2.6 \times 10^{11} 
\left(\frac{\epsilon}{\ev}\right)^{-1}\ev.
\label{PP_threshold}
\ee

The spectrum of electrons generated by pair production in the interaction 
of two photons with energy $E_\gamma$ and $\epsilon_\gamma$ with  
$E_\gamma \gg \epsilon_\gamma$ is \cite{PP83}:
\begin{multline}
\dd{N_e}{E^\prime_e}\left(E_\gamma, \epsilon, E^\prime_e\right)=\sigma_T
\frac{3}{64}\frac{1}{\epsilon^2 E_\gamma^3} \left[
\frac{4A}{ (A-E_e^\prime)E_e^\prime}\log 
\left(4\epsilon(A-E_e^\prime)\frac{E_e^\prime}{A}\right) 
-8\epsilon A+ \right. \\
\left.
2(2\epsilon A-1)\frac{A^2}{(A-E_e^\prime)E_e^\prime}+
-\left(1-\frac{1}{\epsilon A}\right)\frac{A^4}{(A-E_e^\prime)^2E_e^{\prime 2}}
\right]
\label{pp}
\end{multline}
where all energies are in units of the electron mass and we put 
$A=\epsilon+E_\gamma$. The range of electron energies in the final
state is defined by:
\be
\frac{A}{2}\left(1-\sqrt{1-\frac{1}{E\epsilon}}\right)\leq 
E^\prime_e\leq\frac{A}{2}\left(1+\sqrt{1-\frac{1}{E\epsilon}}\right).
\ee

\subsubsection{Inverse Compton Scattering}
The cross section for ICS ($e \gamma_b \rightarrow e \gamma$) is given
by the well-known Klein-Nishina expression (e.g.\cite{R-L,lee98}):
\begin{equation}
\sigma_{\rm ICS} = \sigma_T \cdot \frac{3}{8} \frac{m_e^2}{s\beta} \left[
\frac{2}{\beta (1+\beta)} (2+2 \beta-\beta^2 - 2 \beta^3) - \frac{1}{\beta^2}
(2-3\beta^2-\beta^3) \ln \frac{1+\beta}{1-\beta} \right],
\label{sigma_ICS}
\end{equation}
where $\beta \equiv (s-m_e^2)/(s+m_e^2)$ is the velocity of the outgoing
electron in the center of mass frame.

As in the case of pair production, we use an approximate formula 
for the spectrum of the produced photons/electrons. 
The approximation holds as long as the incoming electron is relativistic. 
The spectrum of electrons resulting from the scattering is 
\cite{blumenthal70,jones68}:
\be
\dd{N_\gamma}{E^\prime_\gamma}\left(E_e, \epsilon, E^\prime_\gamma\right) = 
\sigma_T
\frac{3}{4\gamma\epsilon}\left[ 2q+\log q+(1+2q)(1-q)
+\frac{1}{2}\left(\Gamma_eq\right)^2\frac{1-q}{1+\Gamma_e q}\right],
\label{ics}
\ee
where $\gamma=\sqrt{1+E_e^2}$, $E_1=E_\gamma^\prime/\gamma$, $\Gamma_e=
4\epsilon \gamma$, $q=\frac{E_1}{\Gamma_e(1-E_1)}$ and the range is 
restricted to $\frac{\epsilon}{E_e}\leq E_1 \leq \frac{\Gamma_e}{1+\Gamma_e}$.
All energies are again in units of the electron mass.

The differential spectrum of the outgoing photons is obtained
by substituting $E_e-E^\prime_e$ to $E^\prime_\gamma$ in
eq.~(\ref{ics}).

By averaging the spectra of equations (\ref{pp}) and (\ref{ics}) over the 
background spectrum we can obtain directly the integrated quantities of 
Eq.~(\ref{boltzmann}):
\be
\int\mathrm{d}\epsilon n(\epsilon) P_{PP} (E_e^\prime ; E_\gamma, \epsilon) =  
\int\mathrm{d}\epsilon n(\epsilon) \dd{N_e}{E^\prime_e}
\left(E_\gamma, \epsilon, E^{\prime}_e\right)
\ee
and
\be
\int\mathrm{d}\epsilon n(\epsilon)P_{ICS, \gamma} 
( E_\gamma^\prime ; E_e, \epsilon) = \int\mathrm{d}\epsilon n(\epsilon) 
\dd{N_\gamma}{E^\prime_e}\left(E_e,\epsilon, E^{\prime}_\gamma \right)
\ee
and similarly for the outgoing electrons.


\subsubsection{The magnetic field}
\label{sec:magnetic}

Synchrotron losses of high energy electrons are introduced as continuous
energy losses. This works perfectly for the range of energies that we
are interested in.

The rate of energy losses can be written as:
\be
\dd{E_e}{t}=-\frac{4}{3}c\sigma_T \frac{B^2}{8\pi}\left(
\frac{E}{m_e c^2}\right)^2 
\braket{\sin\alpha} \simeq 7.9\times 10^{-17} \braket{\sin\alpha}
\left( \frac{B}{\mathrm{nG}} 
\right)^2 \left( \frac{E}{\gev} \right)^2 \frac{\gev}{\mathrm{year}}
\ee
where $\alpha$ is a pitch angle. 

The spectrum of photons radiated by synchrotron emission by an electron with 
energy $E$, assuming isotropy of the electron distribution, is 
\cite{R-L}
\be
\dd{P}{E'}=\frac{2\sqrt{3}}{3}\mu_B B \alpha \hbar F\left( 
\frac{E'}{E_C} \right)
\ee
where $\mu_B=9.27\times 10^{-21}$~erg/G is the Bohr magneton, $\alpha$ is 
the fine structure constant, 
\be
E_C=3\mu_B B \gamma^2 \simeq 6.6 \times 10^{-20} \frac{B}{\mathrm{nG}}
\left( \frac{E}{\gev} \right)^2 \gev
\ee
is the critical energy and 
\be
F(x)=x\int_x^\infty K_{5/3}(\xi)\mathrm{d}\xi
\ee
with $ K_{5/3}(\xi)$ the modified Bessel function of fractional order. The 
power spectrum peaks at $E_\gamma \simeq 0.29 E_c$. 

\section{Results}
\label{sec:results}

In this section we discuss the results of our numerical calculations
of the development of the electromagnetic cascade from nearby sources of 
UHECRs. We first discuss the case in which the cascade develops in the 
absence of magnetic field in the intergalactic medium (\S \ref{sec:withoutB}). 
In \S \ref{sec:withB} we apply our calculations to a magnetized universe.

\subsection{The unmagnetized case}
\label{sec:withoutB}

Numerical calculations of the propagation of UHECRs from 
point sources appear to give small scale anisotropies compatible 
with the observations carried out with the AGASA experiment for
a source density of $10^{-6}-10^{-5}~\rm Mpc^{-3}$, corresponding
to a source luminosity above $10^{19}$ eV of $2\times 10^{43}-2\times
10^{42}\rm erg \, s^{-1}$. The best fit injection spectrum is $E^{-2.6}$
for sources with no luminosity evolution and $E^{-2.4}$ for sources
with strong luminosity evolution, $L(z)\propto (1+z)^4$. 
We followed the propagation of UHECRs 
with the simulation used in \cite{demarco1,blasi2003} in which the
interactions are implemented with the SOPHIA package. At each 
step of the simulation, the spectra of gamma rays (from the decay
of neutral pions) and of electrons (from the decay of charged
pions) are determined and used as source functions for the gamma
rays and electrons respectively in the coupled Boltzmann equations 
for the development of the electromagnetic cascade. 

\begin{figure}
  \begin{center}
    \epsfig{figure=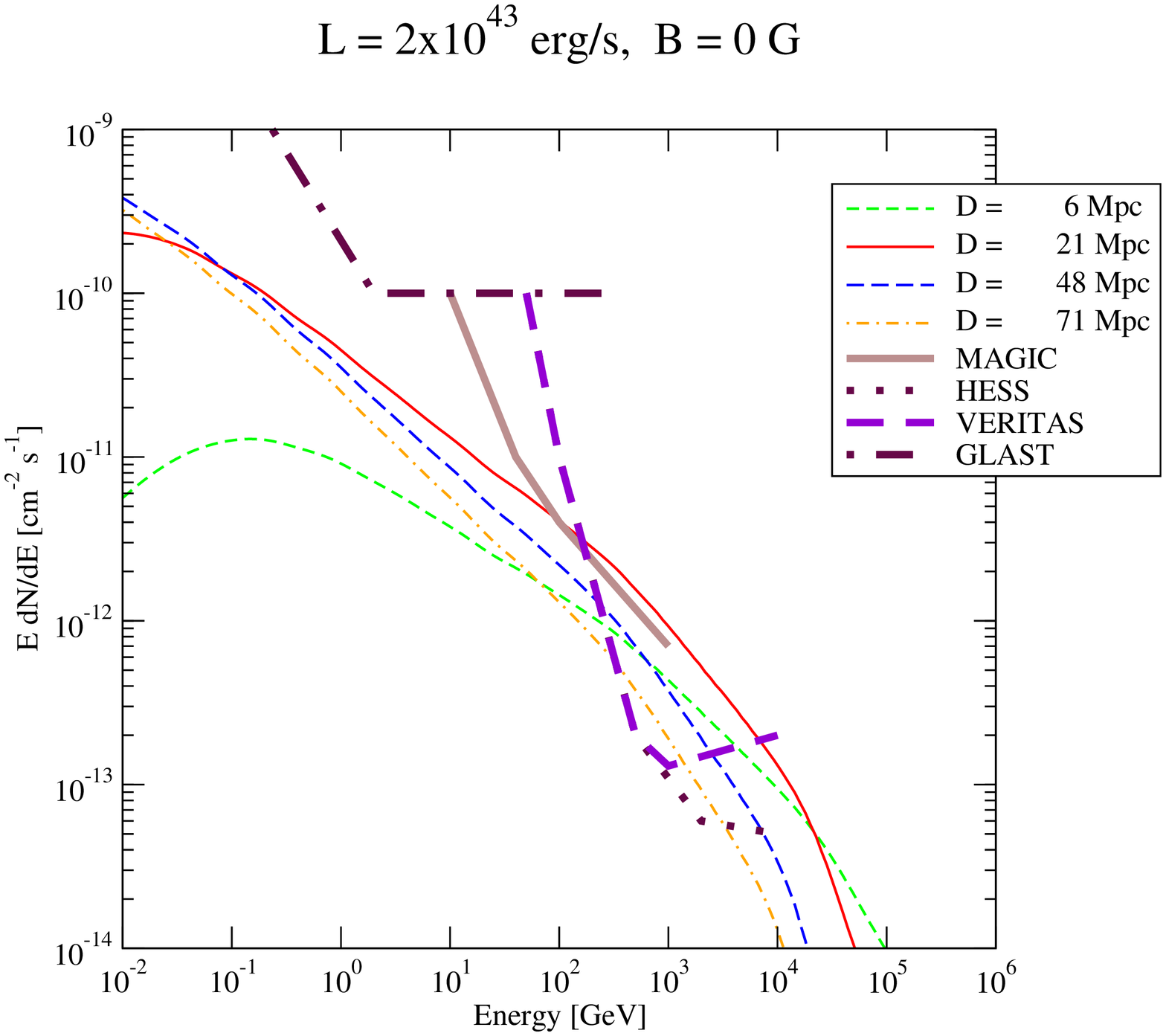,width=5.8cm, angle=-90}
	\hfill
	\epsfig{figure=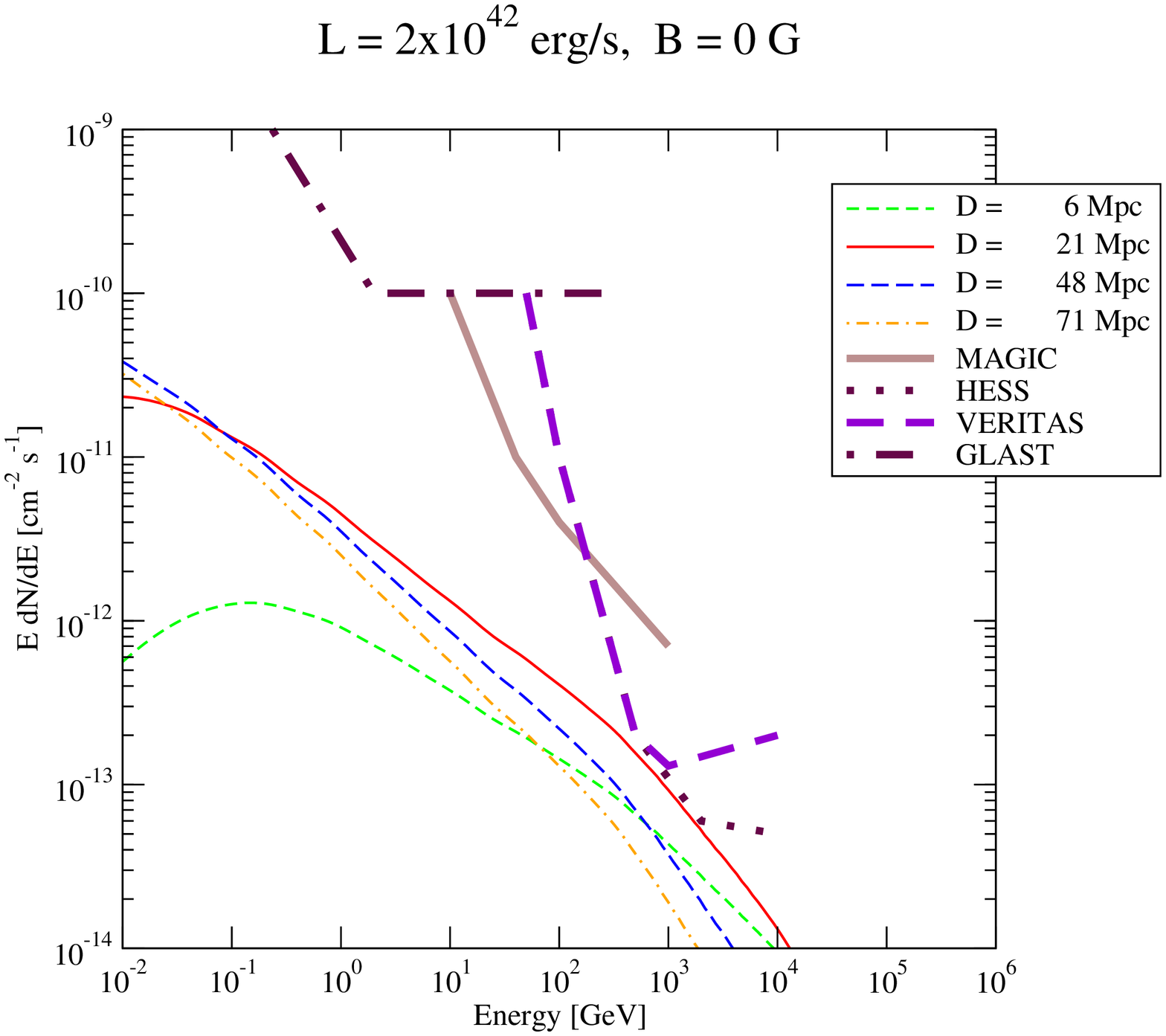,width=5.8cm, angle=-90}
   	\caption{Flux of gamma ray at the Earth as resulting from the 
injection of cosmic rays in continuous point sources located at the 
distances indicated in the graph. The thick lines are the sensitivities
of MAGIC, HESS, VERITAS and GLAST. The left (right) panel is for a source 
with luminosity $2\times 10^{43}~\rm erg \, s^{-1}$ ($2\times 10^{42}~
\rm erg \, s^{-1}$) at energies larger than $10^{19}$ eV.}
\label{fig:fluxB0}
\end{center}
\end{figure}

The results of our calculations for a proton injection spectrum $E^{-2.6}$
are shown in Fig. \ref{fig:fluxB0}
for a source located at 6 Mpc (short dashed line), 21 Mpc (solid
line), 48 Mpc (long dashed line) and 71 Mpc (dash-dotted line). 
In the same plot we show the sensitivities of MAGIC, HESS, VERITAS 
and GLAST (thick lines). 

For the case with higher luminosity (left panel) the predicted fluxes 
should be detectable by the experiments HESS and VERITAS for all the
considered distances. MAGIC should detect a signal only if the 
source is at about 20 Mpc distance. The corresponding fluxes for the 
case with source luminosity $2\times 10^{42}~\rm erg s^{-1}$ are 
slightly below the sensitivities of these telescopes. None of the 
predicted fluxes appears to be detectable in the lower energy range, 
accessible to GLAST. It is worth to remind that the largest distance
considered in the calculations is comparable with the loss length
of UHECRs with energy in excess of $10^{20}$ eV, for which a counterpart 
is still lacking. 

The flux of gamma rays does not necessarily 
decrease for larger distances to the source. This is due to the fact that 
for too small distances, the electromagnetic cascade has not enough time
to develop and the lower energy part of the spectrum does not reach the
equilibrium shape. This is the reason why in Fig. \ref{fig:fluxB0} 
the spectrum of gamma rays from a source at distance 6 Mpc (short 
dashed line) for a wide range of energies is below that corresponding 
to a source at larger distances.
 
Another point to keep in mind is that the calculation in \cite{blasi2003}
were carried out for a class of sources with identical luminosity. It
would probably be more realistic to imagine that the sources of UHECRs
have some kind of luminosity function, and that the value used here is
some sort of average luminosity. In this case it is quite plausible 
that sources with cosmic ray luminosity larger than 
$2\times 10^{43}~\rm erg~ s^{-1}$ exist in nature. 
Since the flux of gamma rays in the cascade scales linearly
with the cosmic ray luminosity, the cascade signal increases correspondingly.

\begin{figure}
  \begin{center}
    \epsfig{figure=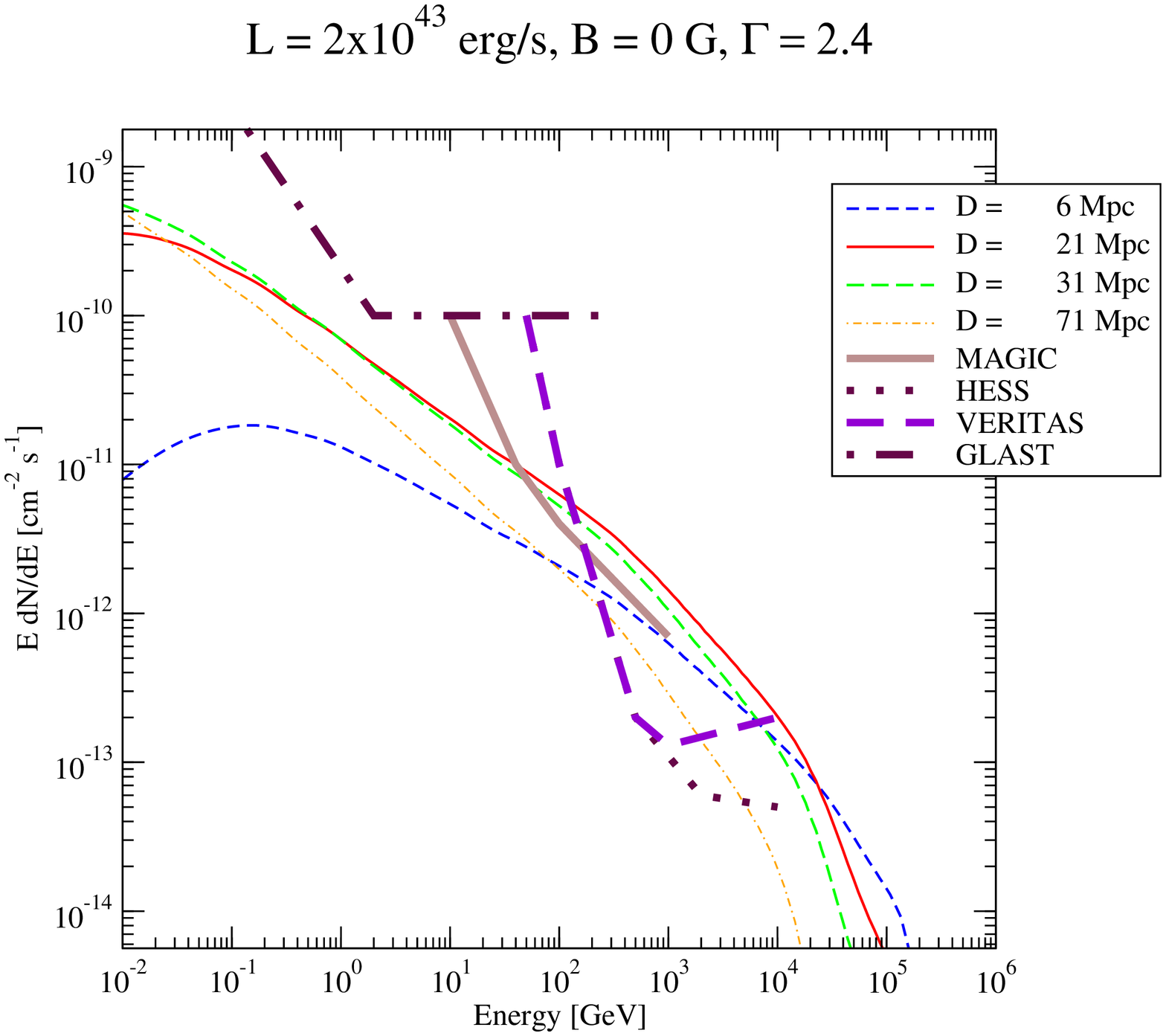 ,width=5.8cm, angle=-90}
	\hfill
	\epsfig{figure=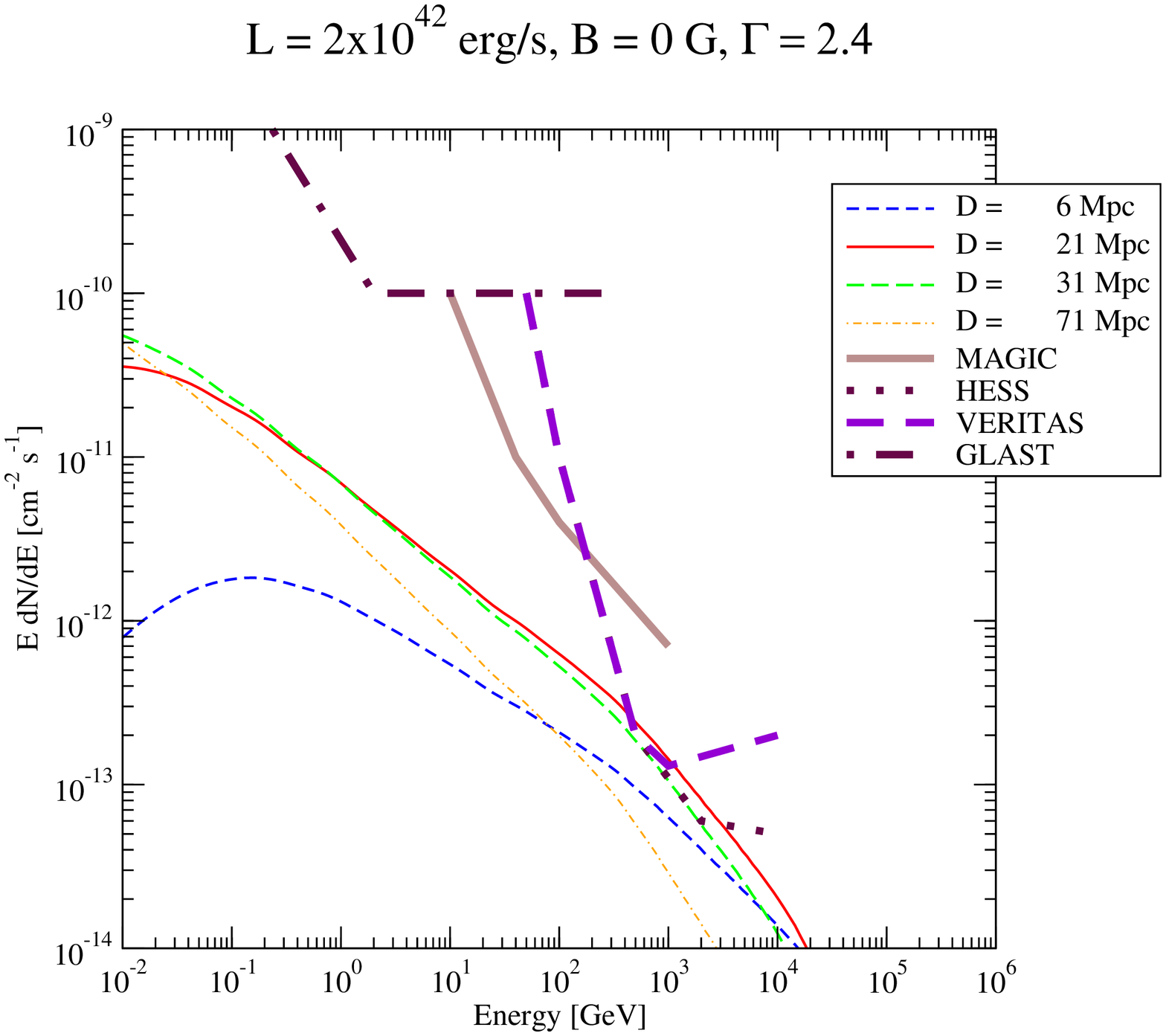 ,width=5.8cm, angle=-90}
   	\caption{Flux of gamma ray at the Earth as resulting from the 
injection of cosmic rays in continuous point sources located at the 
distances indicated in the graph. The thick lines are the sensitivities
of MAGIC, HESS, VERITAS and GLAST. The left (right) panel is for a source 
with luminosity $2\times 10^{43}~\rm erg \,s^{-1}$ ($2\times 10^{42}~\rm erg \,s^{-1}$)
at energies larger than $10^{19}$ eV. The luminosity 
evolution of the sources is taken as $L(z)\propto (1+z)^4$ and the 
injection spectrum of UHECRs as $E^{-2.4}$.}
\label{fig:fluxB0evol}
\end{center}
\end{figure}

For sources with a strong luminosity evolution, the flatter required injection
spectrum in the form of UHECRs slightly increases the predicted cascade
fluxes, as shown in Fig. \ref{fig:fluxB0evol}.

\subsection{The case of a magnetized universe}
\label{sec:withB}

The development of electromagnetic cascades in the universe is very 
sensitive to the presence of magnetic fields, mainly in two ways: 1)
the presence of a magnetic field modifies the duration of the cascading 
processes as observed at the Earth, due to the deflection of the charged
components in the cascade; 2) electrons with very high energy can lose
energy by synchrotron emission, therefore subtracting energy from the 
cascade. 

The first effect is important only for the case of bursting sources. In
that case the duration of the burst in the form of the electromagnetic
cascade may depend on energy, namely the duration of the burst appears
to be different at different energies. Very tight limits on the magnetic
field in the IGM can be imposed in principle from the detection or non
detection of the lowest energy part of the cascade \cite{dai2002}. Since 
we consider here continuous sources, we will not investigate any further this
first effect of the magnetic field, and we will concentrate our attention
on the role of synchrotron losses of the electron component. 

At very high energies, the cross section for ICS is in the Klein-Nishina 
regime. This is the energy region where it is easier for the synchrotron
losses to dominate over ICS. Therefore we expect that the role of the 
magnetic field is particularly relevant when the field is concentrated
around the source, where the electrons in the cascade have the highest
energies. Moreover, as discussed in \S \ref{sec:magnetic}, synchrotron
losses may become dominant on ICS in the energy region around $10^{20}$
eV if the field is larger than $\sim 10^{-10}$ G. 

\begin{figure}
  \begin{center}
    \epsfig{figure=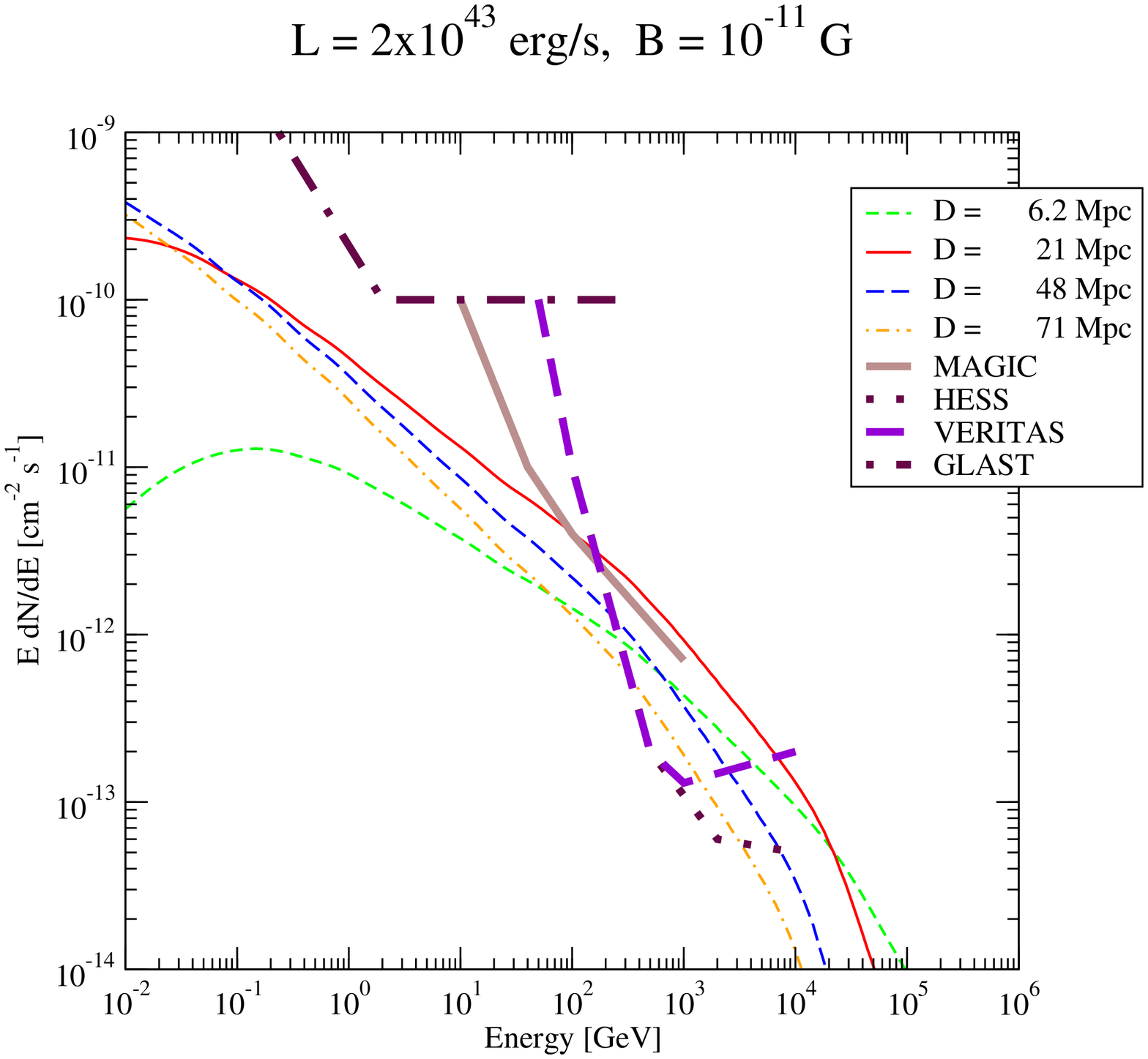,width=5.8cm, angle=-90}
   	\hfill
    \epsfig{figure=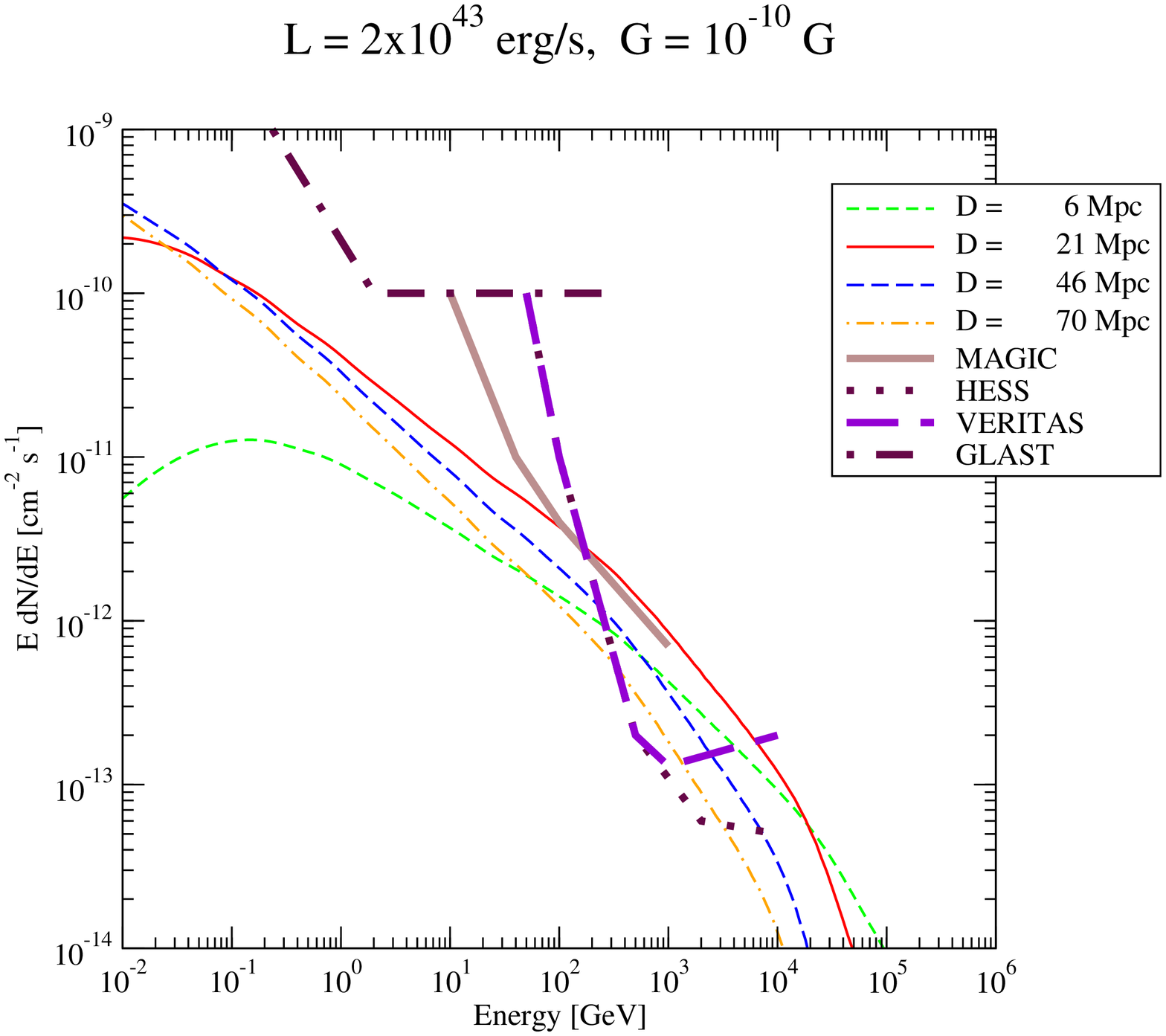,width=5.8cm, angle=-90}
  
    \epsfig{figure=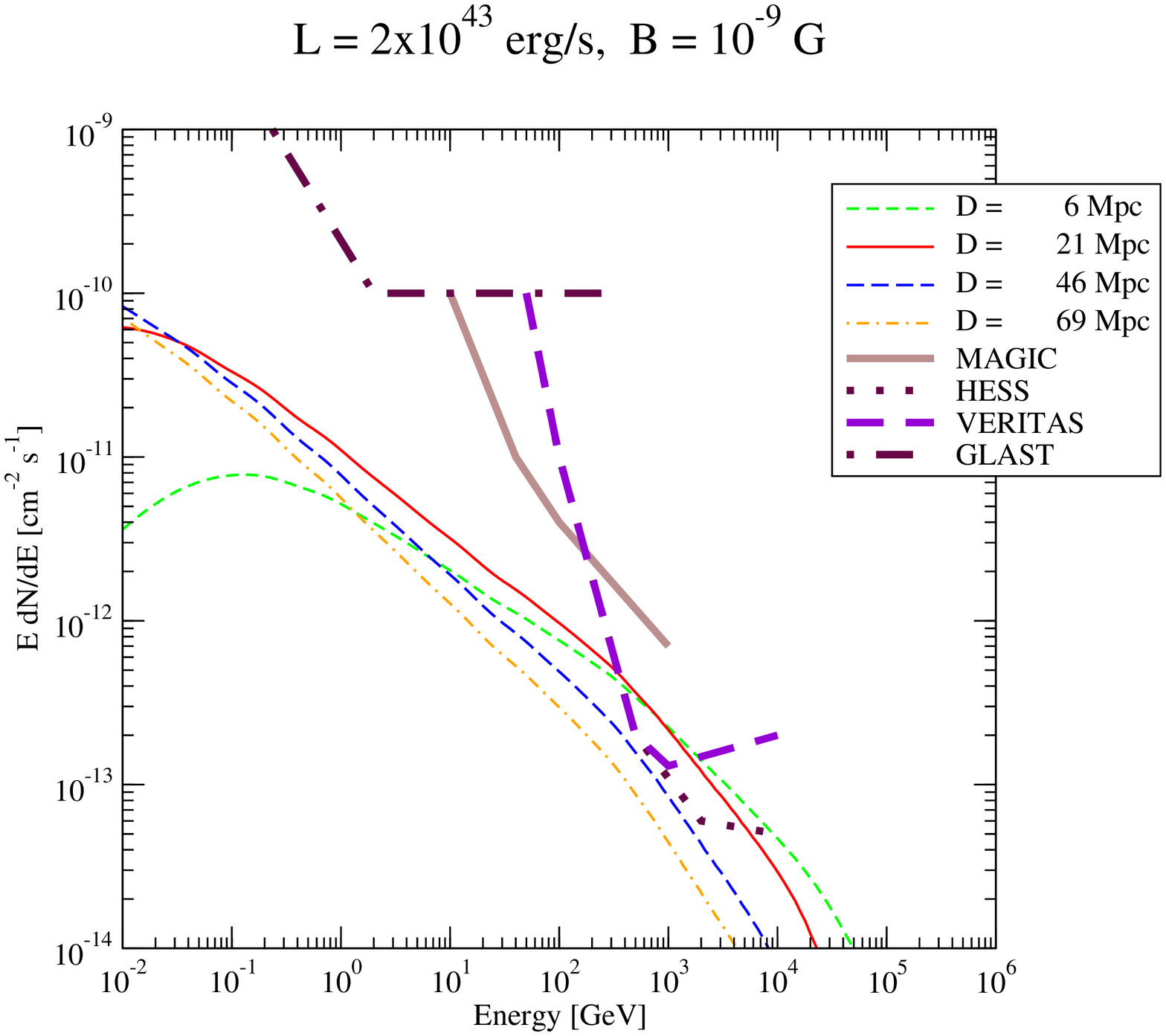,width=5.8cm, angle=-90}
   	\hfill
    \epsfig{figure=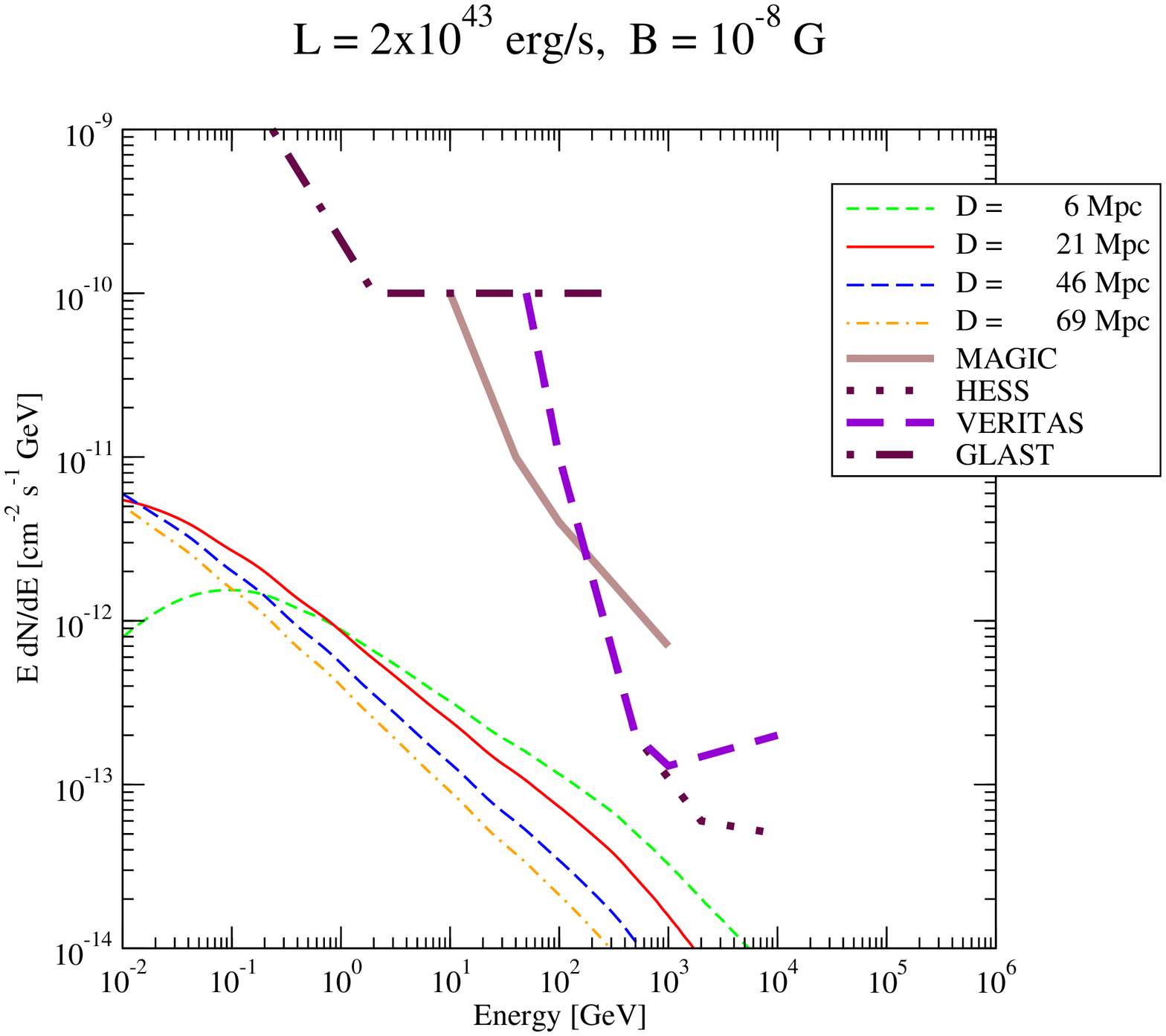,width=5.8cm, angle=-90}
   	\caption{Flux of gamma rays at the Earth for a magnetic field
in the intergalactic medium $B=10^{-11}, 10^{-10}, 10^{-9}, 10^{-8}$~Gauss 
and a source with luminosity $2\times 10^{43}~\rm erg \, s^{-1}$ 
at energies larger than $10^{19}$ eV.}
\label{fig:flux43B001_10}
\end{center}
\end{figure}

Our results for the case of a source with luminosity $2\times 10^{43}~
\rm erg~s^{-1}$ are summarized in Fig. \ref{fig:flux43B001_10} for
four values of the magnetic field, namely $10^{-11}$ G, $10^{-10}$ 
G, $10^{-9}$ G and $10^{-8}$ G. 
The distances to the source are taken 
as in Fig. \ref{fig:fluxB0}, while the source luminosity in the form 
of cosmic rays with energy above $10^{19}$ eV has been taken to be 
$2 \times 10^{43}~\rm erg~s^{-1}$.

The effect of the magnetic field on the development of the electromagnetic
cascade is clear: for magnetic fields smaller than $10^{-9}$ G the predicted
fluxes start to be at the level of sensitivity of VERITAS and HESS, and
are undetectable for larger magnetic field strengths. On the other hand,
one has to keep in mind that a magnetic field of $10^{-9}$~G widespread
in the whole IGM is already the highest possible value compatible with Faraday
rotation measurements \cite{burles}. 
Larger fields are possible only on more compact 
regions: for this reason we also explore the situation in which a magnetic
field is present only in a neighborhood of the source or near the Earth.
A recent investigation of the effects of this magnetic field on the 
propagation of UHECRs has been presented in \cite{grasso,ensslin}. 
In particular in Ref. \cite{grasso} it was shown that the average magnetic 
fields are quite smaller that $10^{-9}$ G, with the exception of the 
filamentary regions where they are of the order of $10^{-9}$ G. Clearly much 
higher fields are achieved in Mpc scale virialized regions such as clusters of 
galaxies. This conclusion does not appear to be confirmed in \cite{ensslin}.

\begin{figure}
  \begin{center}
   \epsfig{figure=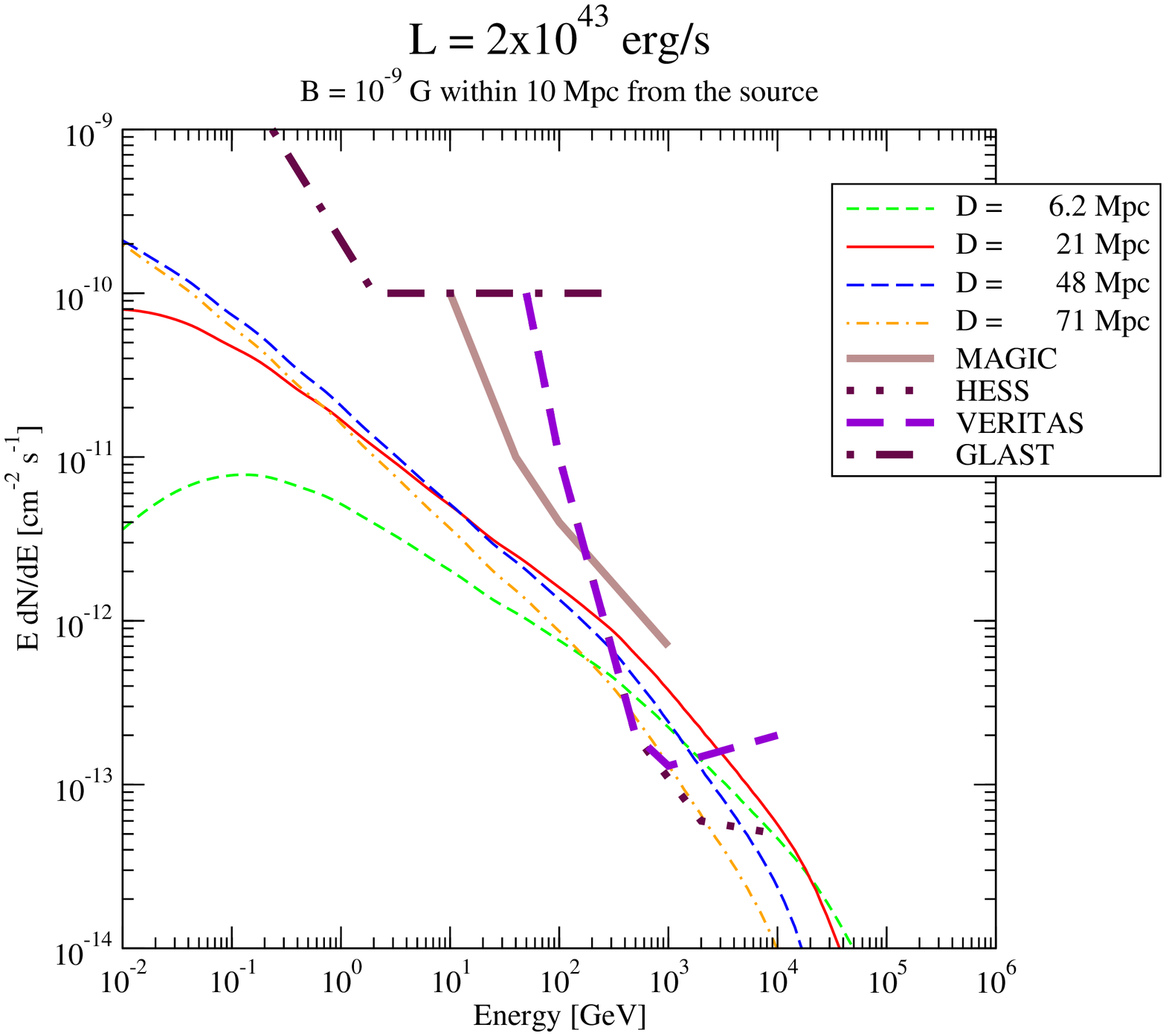,width=5.8cm, angle=-90}
	\hfill 
   \epsfig{figure=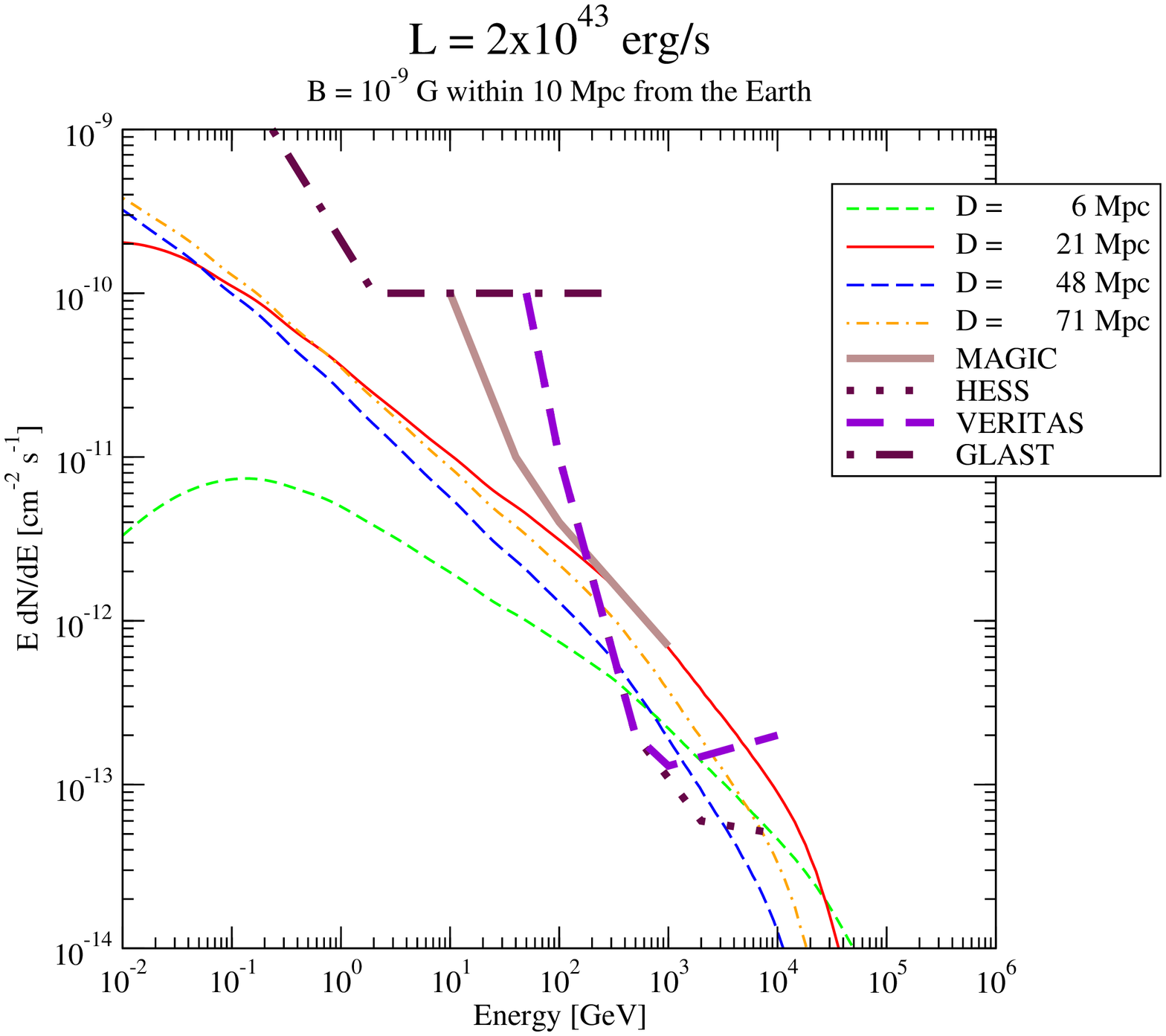,width=5.8cm, angle=-90}
   	\caption{Flux of gamma rays at the Earth for a magnetic field
in the intergalactic medium $B=10^{-9}$ Gauss located in a region of 
10 Mpc around the source and the Earth respectively. The source luminosity is $2\times 10^{43}~\rm erg \, s^{-1}$ at energies larger 
than $10^{19}$ eV.}
\label{fig:flux43nearSE}
\end{center}
\end{figure}

In the left panel of Fig.~\ref{fig:flux43nearSE} we plot the fluxes of 
gamma rays calculated if the source is located in a region of 10 Mpc where 
the magnetic field is $10^{-9}$ G. 

Similarly, in the right panel of Fig.~\ref{fig:flux43nearSE} we plot the 
fluxes of gamma rays calculated in the case that the Earth is located in 
a region of 10 Mpc in which the field is $10^{-9}$ G. This might be the 
case for the Local Supercluster.

It is easy to see that if the magnetic field is concentrated close to the
detector and not close to the source, its effect on the development
of the cascade is negligible, as can be inferred by comparing 
the right panel of Fig.~\ref{fig:flux43nearSE} 
and the left panel of Fig.~\ref{fig:fluxB0}.

\begin{figure}
  \begin{center}
    \epsfig{figure=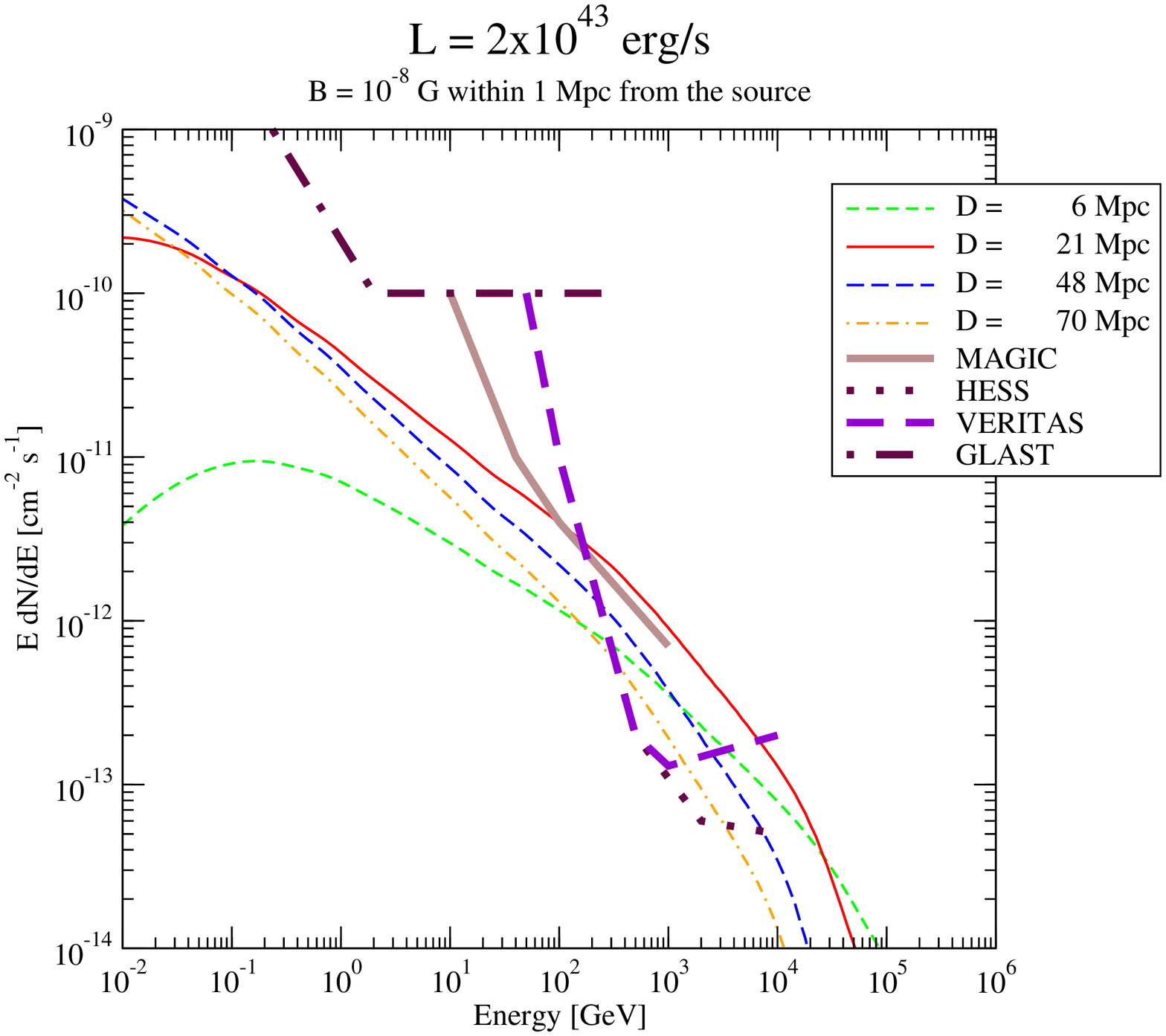,width=5.8cm, angle=-90}
  	\hfill
    \epsfig{figure=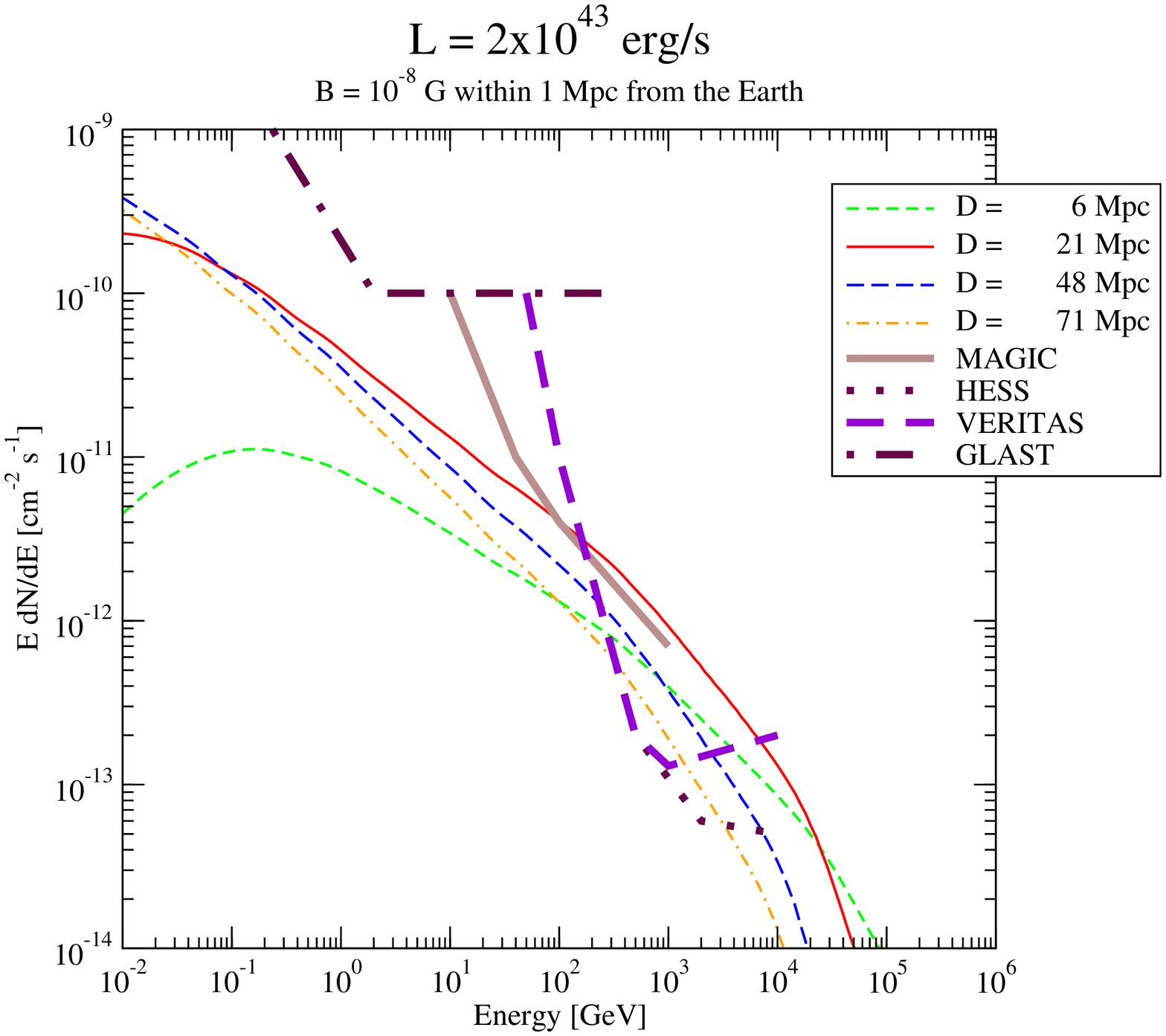,width=5.8cm, angle=-90}
   	\caption{Flux of gamma rays at the Earth for a magnetic field
in the intergalactic medium $B=10^{-8}$ Gauss located in a region of 
1~Mpc around the source and around the Earth respectively. The source 
luminosity is $2\times 10^{43}~\rm erg \, s^{-1}$ 
at energies larger than $10^{19}$ eV.}
\label{fig:fluxNSE1Mpc}
\end{center}
\end{figure}

In the left (right) panel of Fig. \ref{fig:fluxNSE1Mpc} we plot the expected
fluxes in the case that the magnetic field is $10^{-8}$~G in a region of
size $1$ Mpc around the source (Earth). The expected
fluxes in these cases are not much affected by the presence of the 
magnetic field: in the {\it near source} region of 1~Mpc almost no reaction
of photopion production occurred, therefore no cascade was developed yet.
In the {\it near Earth} region the cascade has already reached in
most cases the final stages, where the effects of synchrotron losses
of electrons are irrelevant compared with ICS.

\section{Conclusions}
\label{sec:conclusions}

The sources of UHECRs are still unknown mainly because
we do not know yet whether they are generated in a few very 
powerful sources or rather in numerous low power sources. This degeneracy
has probably been broken after the discovery of small scale anisotropies
in the arrival directions of UHECRs in the AGASA experiment, although 
the statistical significance of this discovery appears to be still 
controversial \cite{finley2003}. In \cite{blasi2003}, the AGASA data were used
to infer that the most likely number density of sources of UHECRs is 
around $10^{-6}-10^{-5}~\rm Mpc^{-3}$, which corresponds to a source
luminosity of $2\times 10^{43}-2\times 10^{42}~\rm erg~s^{-1}$ at 
energies above $10^{19}$ eV. These estimates, obtained in the absence 
of intergalactic magnetic fields, remain valid as long as the magnetic
field does not exceed $\sim 10^{-10}$ G over cosmological distances. 

During their propagation from the sources to the Earth UHECRs suffer
photopion production if their energy is above threshold for this process, 
therefore generating charged and neutral pions. While the former decay 
mainly into electrons and neutrinos, the latter generate gamma rays. 
Both gamma rays and electrons initiate an electromagnetic cascade that
transfers energy from high energies to lower energies, more easily 
accessible to gamma ray telescopes. We calculated the development of this
electromagnetic cascade with and without the presence of an intergalactic
magnetic field, and investigated the detectability of the gamma ray 
signal with ground based telescopes such as MAGIC, VERITAS and HESS, as
well as with space-borne telescopes such as GLAST \footnote{The fluxes
derived in the present paper could be higher by up to a factor of $\sim 2$ 
due to the contribution from proton pair production, that was neglected here.}.

The cascade radiation appears to be detectable by VERITAS and HESS for 
a source luminosity around $2\times 10^{43}~\rm erg~s^{-1}$ in the form
of cosmic rays with energy larger than $10^{19}$ eV, if the magnetic field 
in the IGM is constant over all the propagation volume and smaller than 
$10^{-9}$ G (This luminosity corresponds to the lower source density 
compatible with the small scale anisotropies observed by AGASA).
If the magnetic field is patched, stronger fields are allowed
close to the Earth without appreciably changing the previous conclusions.
If the region with stronger fields is close to the source location, the
initial stages of the cascade may be affected if the magnetized region is 
large enough, so that the energy content of the
cascade gets suppressed. A large field close to the Earth does not affect
the conclusions in any appreciable way, since the cascade is already fully 
developed at that point. A magnetic field of 1nG near the source implies
slightly lower gamma ray fluxes if it is extended in a region of about
10 Mpc. A stronger field, of say 10 nG, in a 1 Mpc region either around 
the source or the observer does not change the results in any appreciable 
way. 

The main uncertainty involved in our calculations is the luminosity of the
sources in the form of UHECRs. The numbers we adopted are derived in 
\cite{blasi2003}, on the basis of the appearance of small scale anisotropies
in the AGASA data and in the assumption of sources that accelerate UHECRs
continuously. For bursting sources, the luminosity per source may be 
many orders of magnitude larger than that adopted here, but it would be 
concentrated within a short time interval. The diffuse and time-spread 
appearance of the observed UHECRs is achieved, within this class of models, 
by assuming the presence of a small magnetic field in the intergalactic medium,
responsible for sizeable time delay that make the burst look like a continuous
signal when observed through charged particles. This is the case for the 
gamma ray burst models \cite{vietri95,waxman95,daniel3} of UHECRs. 

A similar time delay is introduced in the development of the electromagnetic 
cascade. Moreover, an additional time delay is introduced due to the presence 
of electrons and positrons in the cascade itself. These effects cannot be 
accounted for within the framework proposed here, and will be considered in a 
forthcoming publication. 


\section*{Acknowledgments}
The work of PB was partially supported through grant COFIN-2002 at Arcetri.
We are grateful to Anita M\"ucke for providing the SOPHIA source code and
for suggestions on its usage.

\bibliographystyle{phaip}    

\end{document}